# Exploiting hidden singularity on the surface of the Poincaré sphere

Jinxing Li[1,†], Aloke Jana[2,†], Yueyi Yuan[1], Kuang Zhang[1]*, Shah Nawaz Burokur[3]*, and Patrice Genevet[2]*

**Affiliations:**

[1] Department of Microwave Engineering, Harbin Institute of Technology, Harbin, 150001, China

[2] Department of Physics, Colorado School of Mines, 1523 Illinois, St, Golden, CO 80401, USA

[3] LEME, Univ Paris Nanterre, 92410 Ville d'Avray, France

*Corresponding author. Email: zhangkuang@hit.edu.cn, sburokur@parisnanterre.fr, patrice.genevet@mines.edu

†These authors contributed equally to this work.

**Abstract:** The classical Pancharatnam-Berry phase, a variant of the geometric phase, arises purely from the modulation of the polarization state of a light beam. Due to its dependence on polarization changes, it cannot be effectively utilized for wavefront shaping in systems that require maintaining a constant (co-polarized) polarization state. Here, we present a novel topologically protected phase modulation mechanism capable of achieving anti-symmetric full $2\pi$ phase shifts with near-unity efficiency for two orthogonal co-polarized channels. Compatible with -but distinct from- the dynamic phase, this approach exploits phase circulation around a hidden singularity on the surface of the Poincaré sphere. We validate this concept in the microwave regime through the implementation of multi-layer metasurfaces. This new phase modulation mechanism expands the design toolbox of flat optics for light modulation beyond conventional techniques.

## Introduction

Geometric phases, originating from the evolution of instantaneous eigenstates of a system's Hamiltonian in parameter space, are universal features that play a fundamental role in a wide range of phenomena. It is a manifestation of the underlying curvature of the parameter space, which was first observed in a generalized interference experiment with vector fields by Pancharatnam [1], later formalized by Berry in the context of adiabatic evolution of quantum mechanical states [2]. Since its discovery, the geometric phase has led down a foundational ground for a variety of ubiquitous features in several areas of physics including topological phase transitions in condensed matter systems, Aharonov-Bohm effect, spin-orbit coupling, polarization manipulation, and singular optics [3-14]. Furthermore, with the rapid advancement of nanophotonic technology, the geometric phase has become a routine tool to manipulate and control wavefront in a spin-dependent manner [15]. In metasurfaces, it specifically accounts for the relative geometric orientation of the meta-atoms and the helicity reversal of evolving polarization states on the Poincaré sphere —popularly known as the Pancharatnam-Berry (P-B) phase [16-25]. However, all existing P-B phase methods fail to modulate the input polarized light, which we define herein as the co-polarized channel.

To address this issue, we propose and experimentally demonstrate a novel phase addressing mechanism that works similarly as the P-B phase but for co-polarized channel, by encircling a C-point polarization singularity in the parameter space. In this approach, we exploit the mapping of the phase dynamics from the eigen parameter space onto the Poincaré sphere, uncovering a hidden phase singularity on the surface of the Poincaré sphere — referred to as the *co-polarized singular phase*. This previously untapped topologically protected singular phase enables the realization of innovative wavefront control devices. The experimental demonstration of the co-polarized singular phase is achieved through manipulating the chiral response of a lossless multi-layer metasurface that operates in the microwave regime. As opposed to the classical P-B phase, the co-polarized singular phase is fundamentally distinct and can be integrated with other phase addressing mechanisms, such as dynamic phase, to achieve asymmetric phase modulation for two orthogonal input polarizations, broadband wavefront engineering and so forth [26], thus enabling advanced wavefront modulation capabilities that go beyond the limitations of conventional techniques.

**Theory**

The polarization properties of paraxial light propagating in a medium, or through any polarization-sensitive device, are described by a two-dimensional incident polarized field, mathematically represented by a ket vector, and a 2×2 matrix representing the polarization response of the media. The change of polarization is therefore mathematically analogous to the evolution of a spin state in a classical two-level quantum system. Each polarization state is uniquely mapped onto the surface of the Poincaré sphere under the Cartesian coordinate of the Stokes vector, where the north and south pole of the sphere correspond to right- and left-hand circular polarization (RHCP and LHCP) respectively, both constituting the orthonormal basis of the polarized light. In this representation, the degree of polarization, that is the amount of light presenting a well-defined polarization state, is given by the distance of the representative polarization point inside the sphere from the center, ranging from zero at the origin for an unpolarized light, to unity (1) at the sphere surface for a completely polarized light. Assuming a Hermitian optical device (lossless) with orthonormal eigenstates $|\boldsymbol{n}_\pm\rangle$ and corresponding eigenvalues $\exp(i\phi_\pm)$, its polarization-dependent Jones matrix can be expressed as:

$$\hat{U}(\hat{n}, \delta) = \exp\left(i\frac{\delta}{2}\hat{n}\cdot\hat{\boldsymbol{\sigma}}^* + i\phi\right) \tag{1}$$

where $\hat{n}$ is the Stokes vector of $|\boldsymbol{n}_+\rangle$ and $\hat{\boldsymbol{\sigma}} = (\hat{\sigma}_x, \hat{\sigma}_y, \hat{\sigma}_z)$ is the vector of Pauli spin matrices (see supplementary text 1). Parameters $\phi$ and $\delta$ correspond to the dynamic phase (also dubbed the propagation phase) and the eigen birefringence, respectively. These quantities are related to the mean and, respectively, the difference between $\phi_+$ and $\phi_-$. Eq. 1 indicates that the polarization evolution for any arbitrary input state is solely determined by the evolution operator $\hat{U}$. Especially, the degeneracy of the Jones matrix ($\hat{U}$) occurs when all three components of the eigen Stokes vector become identically zero — corresponding to a completely unpolarized eigenstate. In other words, the degeneracy point lies exactly at the center of the Poincaré sphere, acting as the source of Berry curvature, namely a virtual magnetic monopole with intensity $-1/2$ gathering magnetic field inward. We should keep in mind that using magnetic parlance for illustration is merely a matter of convention. As shown in Fig. 1A, the magnetic flux enclosed by the loop associated with a polarization evolution along a closed curve represents the well-known P-B phase in optics. For instance, when an RHCP state $|R\rangle$ is converted to LHCP state $|L\rangle$ by a device with eigenstate $|n\rangle$ whose Stokes vector has an azimuth $2\psi$, state $|R\rangle$ evolves from north pole to south pole along the meridian at $2\psi - \pi/2$ (blue meridian). Compared to a device with an azimuth zero, evolving along the black dashed meridian at $-\pi/2$, the solid angle swept counterclockwise from the black meridian to the blue meridian is $\Omega_+ = 4\psi$ (blue area). Conversely, input $|L\rangle$ is converted to $|R\rangle$ along the same meridian with an opposite sweeping direction of the solid angle, and thus $\Omega_- = 4\pi - \Omega_+$. The magnetic flux is equal to half of the area, i.e. $\gamma_+ = -\gamma_- = -2\psi$, where the sign on the two orthogonal cross-polarized channels is always opposite. Continuously changing the azimuth $\psi$ of the device's eigenstate up to $\pi$, a linearly varying P-B phase accumulation as a function of $\psi$, ranging from 0 to $2\pi$, can be imparted on the cross-polarized wave.

Meanwhile, the dynamic phase is generic and has been widely applied for phase engineering. Such phase corresponds to the mean phase $\phi$ of the eigenvalues $\exp(i\phi_\pm)$, which is equally applied to both eigen polarization channels. Notably, the dynamical phase can be set to zero by selecting opposite eigenvalues, i.e. $\phi_+ = -\phi_-$. Since it is trivial and exhibits homogeneously for any polarization state, the dynamic phase is monotonically distributed on the Poincaré sphere as shown in Fig. 1B.

Considering the polarization properties of both the P-B and dynamic phases, it is evident that a non-trivial phase for the co-polarized wave is still missing. The transmission amplitude in a pure co-polarized channel drops to zero at the antipodal point, directly opposite the incident polarization on the Poincaré sphere. This basic observation reveals that a singularity exists only when considering a projection onto the co-polarized transmission channel. A continuous and topologically protected $2\pi$ phase shift accumulation is expected when choosing a continuous parametric evolution that traces a polarization trajectory around the singularity.

Consider the input is a RHCP state, denoted as $|R\rangle$, and the polarization evolution on the Poincaré sphere is governed by the Jones matrix $\widehat{U}(\hat{n}, \delta)$, characterized by its eigen birefringence $\delta$ and eigen ellipticity $\chi$. The input state precesses counterclockwise around the eigen axis $\hat{n}(\chi)$ by an amount of $\delta$, tracing the red curve from $|R\rangle$ to $|\Psi\rangle$ as shown in Fig.1C. The projection of $|\Psi\rangle$ back to the initial input state $|R\rangle$, is followed by the blue dashed curve representing the shortest geodesic, which can be mathematically performed with the projection operator as $\widehat{P}_+|\Psi\rangle = |R\rangle\langle R||\Psi\rangle$. Notably, the initial and projected states are exactly at the same position on the Poincaré sphere, i.e. the north pole. However, the final projected state picked up an additional phase which can be distinctly seen from the relative configuration of the initial and final polarization state $|R\rangle$. This additional phase arises due to the underlying curvature of the trajectory on the Poincaré sphere. Importantly, this phase accumulation is independent of the rate at which the path is traversed; rather, it depends solely on the geometry of the Stokes parameter space. Therefore, the projected phase in the co-polarized channel is purely geometric in origin, which can be referred to as the co-polarized geometric phase. The complex co-polarized transmittance across eigen parametric coordinates $\mathbf{R} = (\delta, \chi)$, can be derived from Eq. 1 as:

$$t_{co} = \cos\frac{\delta}{2} + i\sin\frac{\delta}{2}\sin 2\chi \tag{2}$$

Now, by tuning the eigen vectors $\hat{n}(\chi)$ while keeping the same azimuth, the orientation of the eigen axis shifts along the meridian. Consequently, the RHCP input $|R\rangle$ follows different trajectories on the Poincaré sphere, leading to varying accumulated geometric phases in the co-polarized channel projective measurement. Interestingly, when the eigen axis of the material crosses the equator $\hat{n}(\chi = 0)$, with a retardance of $\delta = \pi$, the RHCP state reverses its helicity. At this point, the accumulated geometric phase upon projection to the initial $|R\rangle$ state depends on the shortest geodesic path followed. A small variation of the eigen parameters around $\delta = \pi$, $\chi = 0$, alters the position of state $|\Psi\rangle$ and thereby modifying the projection path. The shortest geodesic would follow any meridian trajectory at the antipodal point, leading to a co-polarized singular phase at the corresponding antipodal point $|L\rangle$ of the input $|R\rangle$ state. This singularity is evident when plotting the phase of the complex projection amplitude $\arg(t_{co})$ at the Stokes parametric position of $|\Psi\rangle$ as shown in Fig. 1C. The proposed co-polarized singular phase represents a distinct variant of the geometric P-B phase, that has a deep topological root in the polarization ellipse dynamics, and its dependence on the system's eigen parameters.

To further analyze the singular behavior of the geometric P-B phase, we plot the polarization ellipse of the intermediate state $|\Psi\rangle$ in the 2D eigen-parameter space in Fig. 2A. As it is very evident that at the center $\delta = \pi, \chi = 0$, a C-point polarization singularity appears with a unity transmission amplitude as we are dealing with only lossless Hermitian systems. Moreover, the state $|\Psi\rangle$ is decomposed into two orthogonal circular components using the respective projection operators: $\hat{P}_+ = |R\rangle\langle R|$ for RHCP and $\hat{P}_- = |L\rangle\langle L|$ for LHCP state. In Fig.2 (C-D), the co-polarized projection reveals the existence of a phase singularity with a vortex like behavior in the eigen parameter space for complex transmission amplitude $t_{co}$. However, it is important to note that the projected co-polarized beam should not be confused with vector vortex or orbital angular momentum (OAM) carrying beams as it always preserves its initial input spatial modes—here a plane wave. This observed co-polarized singular phase provides a foundation for designing the meta-atoms to modulate wavefronts in a spin-preserved manner. The transmission amplitude of the co-polarized component is zero at the singular C-point and reaches maximum unity at the edges of the parameter space as shown in Fig. 2C. Moreover, three closed loops in the eigen parameter space—the solid (outside), dashed (crossing) and dotted curve (enclosing the singular C-point) are considered. The corresponding trajectories of the state $|\Psi\rangle$, along with the co-polarized phase distribution, are shown on the surface of the Poincaré sphere (Fig. 2B). When the singular C-point is enclosed in the eigen parameter space such as along the dotted trajectory, the co-polarized transmission channel undergoes a complete 0 to $2\pi$ phase modulation, which validates the proposed concept. In contrast, the cross-polarized channel has a constant phase across the eigen spectrum as shown in Fig. 2F, and a maximum transmission amplitude at the C-point, gradually decays toward the edges. Considering any arbitrary path enclosing this singularity, that is assuming a path in material parameter space that enables continuously changing the $(\delta, \chi)$ values to circulate around the singular point $(\delta = \pi, \chi = 0)$, a $2\pi$ phase accumulation can be obtained. This parametric circulation of the material parameters can potentially be realized, for example, by tailoring the structural properties of a set of meta-atoms.

**Results and Discussion**

In addition to $2\pi$ phase coverage, the amplitude is also crucial since it delimits the transmission efficiency in applications, which can be derived as $\cos(\pi/4-\chi)$. To explicitly display the topological relationship between the complex amplitude of the path around the singularity, the sphere in Fig. 1C is expanded according to the amplitude and phase at each point, as shown in Fig. 3A. The conical surface denotes the transmission amplitude, and the tip corresponds to the singularity at the south pole of the sphere, indicating as well that the phase projected on the bottom plane gradually varies by $2\pi$ around the singularity. Significantly, the path on the top of the conical surface offers a constant transmission coefficient approaching unity, corresponding to the case of optical rotation with $\chi = \pm\pi/4$, i.e., the eigenstate of the meta-atom and incident polarization are consistent (both are CP here), and the co-polarized singular phase imposed to the RHCP preserved transmittance is linear to $\delta/2$. Exploiting circular birefringent meta-atoms, that is designing nanostructures that exhibit polarization rotation by manipulating the phase delay of $|L\rangle$ and $|R\rangle$ using optical chirality, we encircle the singularity by eight representative elements. As shown in Fig. 3B, the meta-atom consists of three layers of copper structure, where both the top and bottom layers contain four identical copper patches forming chiral $C_{4z}$ symmetry to protect the eigenstate as CP. The corresponding top and bottom patches are electrically connected by vertical metal vias, carefully assembled to avoid touching the middle ground plane. Due to this structural configuration, the meta-atom can manipulate circular birefringence $\delta$ quasi-linearly by the relative rotation angle $\theta$ between the top and bottom patches, which follows $\delta \approx 2\theta$ and the phase of the two co-polarized channels is $\zeta_+ = -\zeta_- \approx \theta$. Elements No. 1 to 4 and No. 5 to 8 correspond to $\zeta_+$ increasing from $-\pi/2$ to $\pi/2$ and decreasing from $\pi/2$ to $-\pi/2$ respectively (see supplementary text S4). Fig. 3C clearly illustrates the spectrum $t_{++}$ of meta-atom No. 3 when $\theta$ varies from $0°$ to $90°$. The amplitude is maintained over 0.9 within 8.8 GHz - 15.5 GHz while $\zeta_+$ is shifted by $\pi/2$. Meanwhile, the corresponding cross-polarized transmission coefficient $t_{-+}$ approaches zero (Fig. 3D), proving the eigenstate of meta-atom is exactly CP and its circular birefringence can be tuned by $\theta$ robustly. Note that the transmission amplitude of $t_{++}$ does not reach unity due to the reflection losses (see Fig. S7).

Metasurface technology exploits the phase shift between the incident and scattered radiation of optical resonators that are dispatched across the interface. The design approach requires spatially distributing the resonators with different geometry to introduce a discontinuous phase profile in agreement with a user-defined wavefront. Here, we provide proof-of-concept wavefront engineering in co-circular polarization by choosing and distributing the meta-atoms to realize phase gradient metasurfaces with an aperture size of 187 mm × 187 mm operating at 10 GHz. The first result exploits the anti-symmetric co-polarized phase depending on the input helicity, refracting $|R\rangle$ and $|L\rangle$ waves at $-30°$ and $+30°$ respectively, as schematically shown in Fig. 4A. The gradually varying phases as a function of the spatial x-position is obtained using numerical solutions representing the meta-atoms' transmission properties, as shown in Fig. 4B, which indeed indicate that both $|R\rangle$ (red) and $|L\rangle$ (blue) co-polarized channels are imposing opposite discontinuous phase values. The experimentally measured normalized far-field patterns presented in Fig. 4C show the anti-symmetric refraction angles, which decrease as a function of the frequency within 8 GHz - 12 GHz. At 10 GHz, the refraction angles of $\theta_{t+}$ and $\theta_{t-}$ are consistent with the ideal anti-symmetric refraction, demonstrating that such an intriguing behavior, identical to the conventional P-B phase, can also occur on the co-polarized channels. As mentioned in various examples in the literature related to the P-B phase, it is possible to further decouple such anti-symmetric response by introducing an additional phase modulation mechanism, such as the dynamic phase. Indeed, by simply tuning the size of the meta-atoms while fixing the relative rotation angle of the structure's patches, the co-polarized singular phase can be preserved. This combination of phases is used to design an asymmetric refractor (Fig. 4D) presenting decoupled and independent phase gradients, indicated as $\Phi_{++}$ and $\Phi_{--}$ in Fig. 4E, for both $|R\rangle$ and $|L\rangle$, respectively. Here, we arbitrarily considered a metasurface design with refraction angle $\theta_{t+} = -15°$ and $\theta_{t-} = 45°$ in the two channels at the operation frequency of 10 GHz. The experimentally measured normalized far-field patterns associated with both channels are depicted in Fig. 4F, showing refraction angles in agreement with the design and exhibiting high transmission efficiency (see supplementary text S5). In addition to the demonstration of anomalous refraction using asymmetric one-dimensional phase gradients, we achieved arbitrary and decoupled co-polarized engineering of complex wavefronts, as often reported using two-dimensional phase profiles, including OAM generation from co-polarized singular geometric phase (see supplementary text S6). Here, the generated OAM does not result from helicity reversal, and consequently the total angular momentum is not conserved in co-polarized transmission channel. OAM is achieved by strategically placing the co-polarized phase elements in a vortex-like spatial phase distribution, and as a result, this process can be utilized to exert torque on the spiral phase plate [27].

**Conclusions**

We introduce and validate a concept of co-polarized singular phase for polarization-dependent wavefront engineering, complementing the classical cross-polarized Pancharatnam-Berry phase. We exploit the full $2\pi$- phase discontinuity originating from a topologically protected singularity, hidden in the parameter space and appearing on the surface of the Poincaré sphere. The reported co-polarized singular phase shares analogy with the reported cross-polarized P-B phase, imposing an opposite phase to any pair of orthogonal polarization channels under co-polarized transmission as well as being compatible with other phase modulation mechanisms. Our work offers new research opportunities for versatile phase engineering beyond conventional manners and provides additional degrees of freedom to manipulate the spin angular momentum and spin-orbit interactions of light.

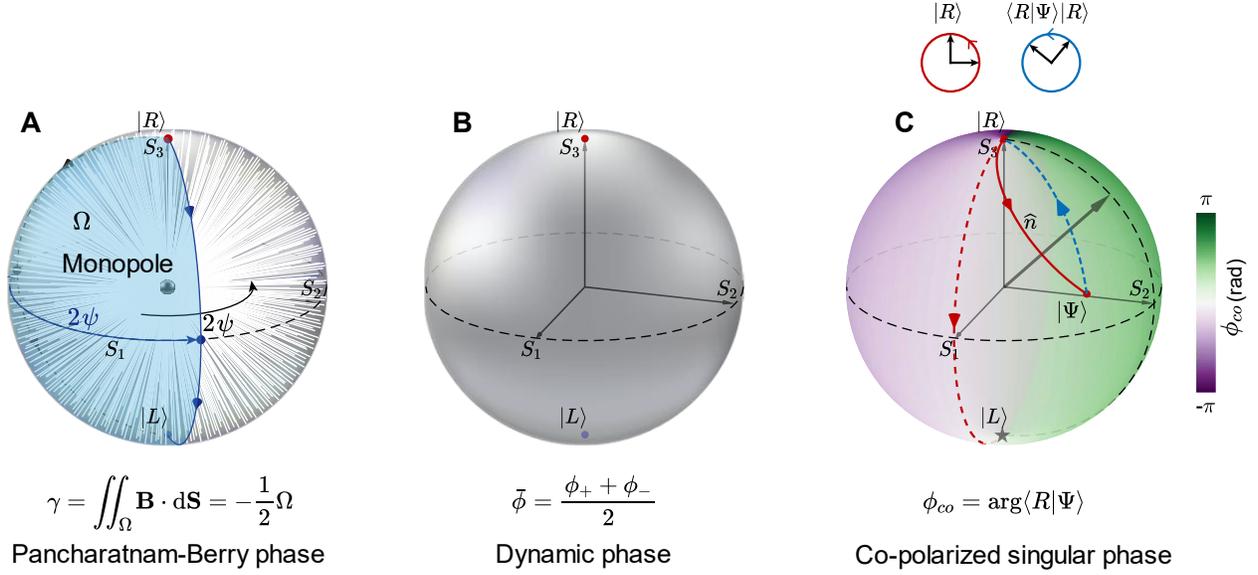

$$\gamma = \iint_\Omega \mathbf{B} \cdot d\mathbf{S} = -\frac{1}{2}\Omega$$

Pancharatnam-Berry phase

$$\bar{\phi} = \frac{\phi_+ + \phi_-}{2}$$

Dynamic phase

$$\phi_{co} = \arg\langle R|\Psi\rangle$$

Co-polarized singular phase

**Fig. 1. Comparison between different phase-addressing mechanisms.** (**A**) Typical Pancharatnam-Berry phase, associated to cross-polarized phase modulation and originating from the degeneracy of the Jones matrix, corresponds to the magnetic flux of a virtual monopole of intensity $-1/2$ positioned at the center of the Poincaré sphere. Considering the transmission of a right incident circularly polarized wave $|R\rangle$ through birefringent materials with eigenstate of azimuth $\psi$ (thick blue arrow), polarization evolves along the meridian from the north to the south pole. Choosing two different meridians, two cross-polarized beams taking two different meridian trajectories would present a phase difference equal to half of the magnetic flux through the blue solid angle $\Omega$ swept counterclockwise between the two meridians (blue and dashed blacklines). (**B**) The dynamic phase is polarization-independent and only depends on mean phase $\bar{\phi} = (\phi_+ + \phi_-)/2$ of the eigenvalues $\exp(i\phi_\pm)$, and therefore applies to any polarization channel equally. (**C**) The evolution of polarization state and the emergence of a co-polarized phase singularity on the Poincaré sphere are governed by the system's eigen parameters $(\delta, \chi)$. The input state $|R\rangle$ undergoes precession around the eigenaxis $\hat{n}(\chi)$ by an angle $\delta$, transitioning to an intermediate state $|\Psi\rangle$, tracing the red curve. The subsequent projection back to the initial state follows the blue dashed curve, acquiring an additional phase, which is evident from the relative configuration of the initial and final $|R\rangle$ states. The accumulated phase, solely depends on the curvature of the traversed path, has a pure geometric origin. Variation in the eigen parameters alters polarization trajectory on the Poincaré sphere and thereby modifying the accumulated geometric phase. When the eigenaxis crosses the equator $\hat{n}(\chi = 0, \psi = constant)$ with $\delta = \pi$, the input $|R\rangle$ state reverses its helicity (red dashed curve), and the accumulated geometric phase through the projective measurement depends on the shortest geodesic path. A very small variation around $(\delta = \pi, \chi = 0)$ shifts the position of $|\Psi\rangle$, affecting the projection path. This leads to a co-polarized singular phase, which lies exactly at the antipodal point $|L\rangle$ denoted with the star mark. This phase singularity emerges in complex amplitude of the co-polarized transmission channel at the Stokes parametric position of $|\Psi\rangle$, representing a topological variant of the geometric P-B phase.

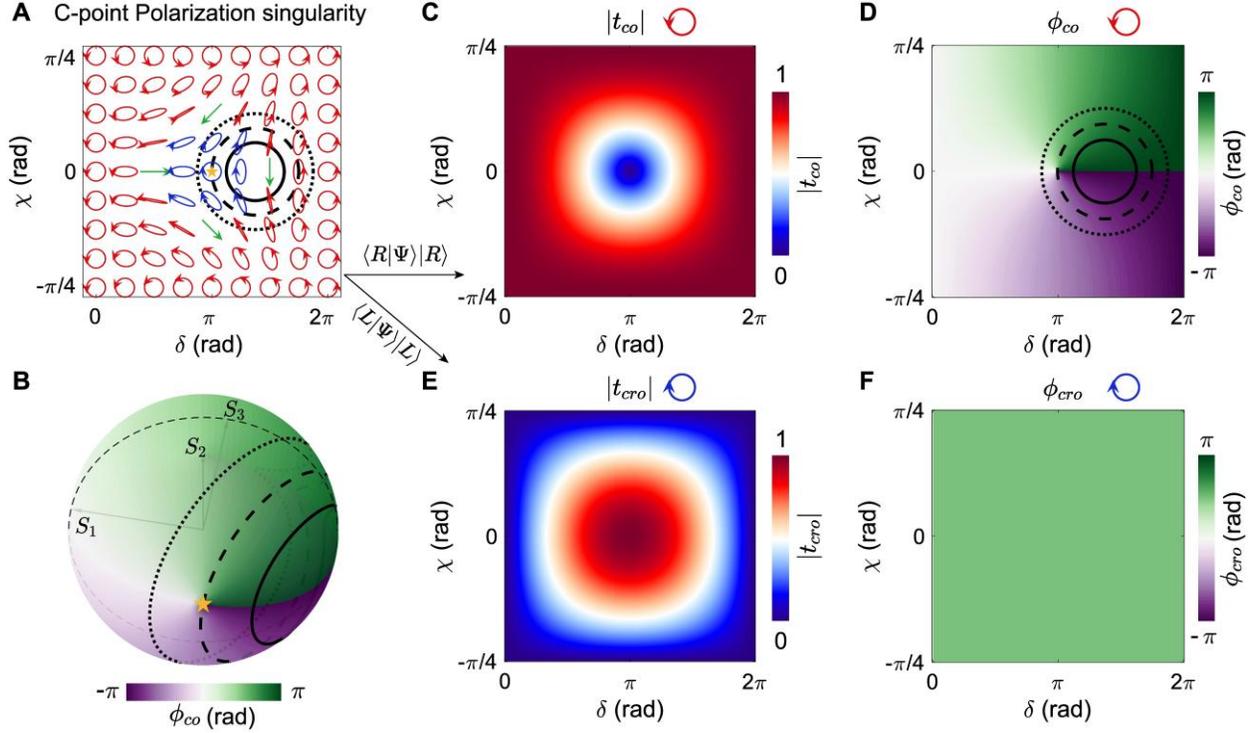

**Fig. 2. C-point polarization singularity and associated co-polarized singular phase in the eigen parameter space** $(\delta, \chi)$. (**A**) Distribution of the polarization ellipses of $|\Psi\rangle$ in the eigen parameter space. The evolution of the input state creates a C-point polarization singularity at $\delta = \pi, \chi = 0$, containing the antipodal LHCP state at the singular point. Three different closed loops outside (solid circle), crossing (dashed circle), and encircling (dotted circle) the singular C-point are considered. (**B**) The corresponding trajectories of $|\Psi\rangle$, along with the co-polarized phase distribution are displayed on the Poincaré sphere. The yellow star represents the position on the Poincaré sphere where the singularity appears. (**C-F**) The decomposition of the $|\Psi\rangle$ into the co- and cross-polarized channels is performed by the respective projection operators, representing amplitude and phase distribution in the eigen-parameter space. Notably, at the C-point, the co-polarized phase becomes singular and circulation of this topologically protected phase enables complete $2\pi$ wavefront modulation in a spin-preserved manner.

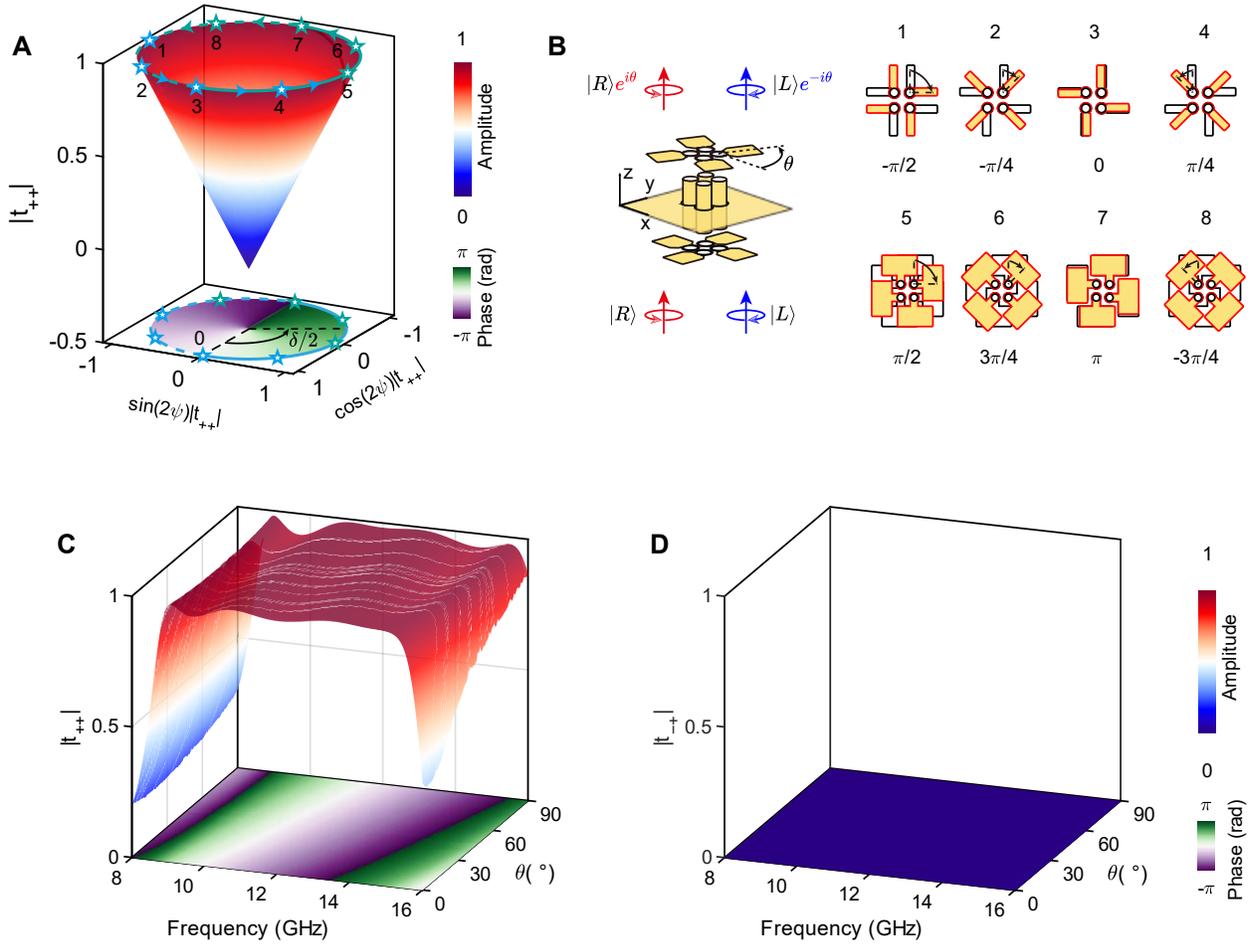

**Fig. 3. Encircling the singularity with unitary transmittance.** (**A**) Expansion of the Poincaré sphere according to the complex co-polarized transmittance. A conical surface representing the amplitude, in addition to the heat map of the projected phase at the bottom, indicates a vanishing amplitude and a rapidly changing phase from [0,2$\pi$] characteristic of phase singularity. Choosing a path that encircles the singular parameter point with unit transmittance can be realized using the designed circular birefringent meta-atoms indicated by stars. (**B**) Schematic of the structure of the circular birefringent meta-atoms and the shapes of the eight elements annotated in (A), designed to tune quasi-linearly ($\delta \approx 2\theta$) the circular birefringence using the elements' relative rotation angle. The (**C**) co- and (**D**) cross-polarized transmittance of an elementary meta-atom with relative rotation angle from 0° to 90°, indicating zero cross-polarized conversion and zero P-B phase.

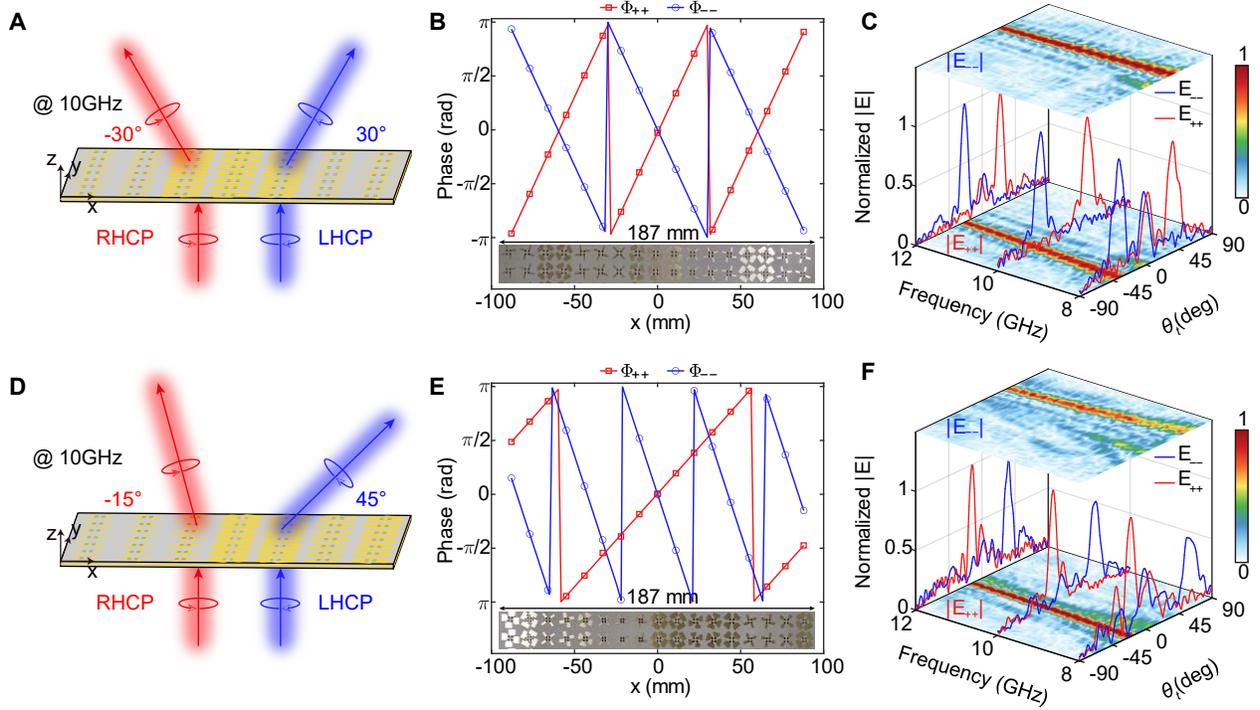

**Fig. 4. Results of metasurfaces for wavefronts tailoring.** (**A**) Schematic of the designed antisymmetric refractor at 10 GHz. (**B**) The designed phase profile of the co-polarized channel of RHCP (red) and LHCP (blue) at 10 GHz of refractor, which are opposite because of the coupling between the responses. The inset is the photograph of the fabricated prototype, which consists of two kinds of structures with different geometric sizes. (**C**) The measured normalized far-field patterns of the two co-polarized channels. (**D**) Schematic of the designed asymmetric refractor at 10 GHz. (**E**) The decoupled phase profile of the two co-channels, and the constituent structures in the inset show multiple geometric sizes since the dynamical phase is introduced. (**F**) The measured normalized far-field patterns of the asymmetric refractor.

**Acknowledgments:**

**Funding:**

National Science Foundation of China grant U23B2014 and 62171165.

**Author contributions:**

Conceptualization: K.Z., P.G.

Methodology: J.L., A.J.

Investigation: J.L., A.J., Y.Y., S.N.B.

Visualization: J.L., A.J., Y.Y.

Writing – original draft: J.L., A.J.

Writing – review & editing: Y.Y., K.Z., S.N.B., P.G.

**Competing interests:** Authors declare that they have no competing interests.

**Data and materials availability:** All data, code, and materials used in the analysis must be available in some form to any researcher for purposes of reproducing or extending the analysis. Include a note explaining any restrictions on materials, such as materials transfer agreements (MTAs). Note accession numbers to any data relating to the paper and deposited in a public database; include a brief description of the data set or model with the number. If all data are in the paper and supplementary materials, include the sentence "All data are available in the main text or the supplementary materials."

**Supplementary Materials**

Materials and Methods

Supplementary Text

Figs. S1 to S14

References (28–31)

# Supplementary Materials for

## Exploiting hidden singularity on the surface of the Poincaré sphere

Jinxing Li[1,†], Aloke Jana[2,†],  Yueyi Yuan[1], Kuang Zhang[1]\*, Shah Nawaz Burokur[3]\*, and Patrice Genevet[2]\*

**Affiliations:**

[1] Department of Microwave Engineering, Harbin Institute of Technology, Harbin, 150001, China

[2] Department of Physics, Colorado School of Mines, 1523 Illinois, St, Golden, CO 80401, USA

[3] LEME, Univ Paris Nanterre, 92410 Ville d'Avray, France

\*Corresponding author. zhangkuang@hit.edu.cn, sburokur@parisnanterre.fr, patrice.genevet@mines.edu

†These authors contributed equally to this work.

**The PDF file includes:**

> Materials and Methods
> Supplementary Text
> Figs. S1 to S14
> References (28-31)

**Supplementary Text**

**Supplementary Note 1: Evolution of Polarization state, Berry curvature and associated geometric phase**

For a lossless Hermitian paraxial optical device with eigenstates $\left|n_{\pm}\right\rangle$ and corresponding eigenvalues $e^{i\phi_{\pm}}$, the polarization-dependent transmission response can be characterized by the governing unitary operator $\hat{U}$, represented by the Jones matrix of the system. The orthonormal polarization bases of the ket vector are considered as right- and left-handed circular polarization (RHCP and LHCP), denoted as $\left|R\right\rangle$ and $\left|L\right\rangle$, respectively. The Stokes parameters corresponding to the direction vector $\hat{n}$ pointing at $(2\psi, 2\chi)$ define the eigenstate of the system (Fig. S1), where $\psi$ represents the azimuthal angle and $\chi$ corresponds to the ellipticity. The eigenstate $\left|n_{+}\right\rangle$ is given by:

$$|n_+\rangle = \begin{pmatrix} S_1 \\ S_2 \\ S_3 \end{pmatrix} = \begin{pmatrix} \cos 2\chi \cos 2\psi \\ \cos 2\chi \sin 2\psi \\ \sin 2\chi \end{pmatrix} \tag{S1}$$

The system's orthonormal eigenvectors $\left|n_{\pm}\right\rangle$ can be expressed in terms of circular polarization basis as follows:

$$|n_+\rangle = \cos\left(\frac{\pi}{4} - \chi\right) e^{i\psi} |R\rangle + \sin\left(\frac{\pi}{4} - \chi\right) e^{-i\psi} |L\rangle = \begin{pmatrix} \cos\left(\frac{\pi}{4} - \chi\right) e^{i\psi} \\ \sin\left(\frac{\pi}{4} - \chi\right) e^{-i\psi} \end{pmatrix} \tag{S2a}$$

$$|n_-\rangle = -\sin\left(\frac{\pi}{4} - \chi\right) e^{i\psi} |R\rangle + \cos\left(\frac{\pi}{4} - \chi\right) e^{-i\psi} |L\rangle = \begin{pmatrix} -\sin\left(\frac{\pi}{4} - \chi\right) e^{i\psi} \\ \cos\left(\frac{\pi}{4} - \chi\right) e^{-i\psi} \end{pmatrix} \tag{S2b}$$

Notably, the eigenvector $\left|n_{-}\right\rangle$ is at the antipodal point of $\left|n_{+}\right\rangle$ on the Poincaré sphere, having an azimuth and ellipticity of $\psi + \pi/2$ and $-\chi$, respectively.

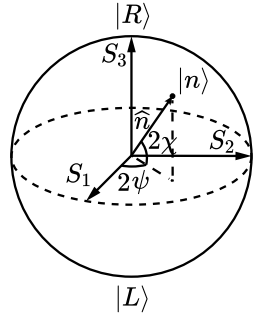

Poincaré sphere

**Fig. S1. Geometrical representation of a polarization state on the Poincaré sphere.** The north and south poles of the Poincaré sphere are $\left|R\right\rangle$ and $\left|L\right\rangle$, and the angular coordinate is $(2\psi, 2\chi)$, where $\psi$ and $\chi$ represent the azimuth angle and ellipticity of the polarization state, respectively.

Any given polarization state can be represented by the intensity normalized Cartesian coordinates of the Stokes vector $(S_1, S_2, S_3)$. The north and south poles of the conventional Poincaré sphere represent the right- and left-handed circular polarization states, respectively, denoted as $\left|R\right\rangle$ and $\left|L\right\rangle$ while the equator corresponds to linearly polarized states. An arbitrary eigen polarization state $\left|n_{+}\right\rangle$ on the surface of the Poincaré sphere is described by angular coordinates $2\psi$ and $2\chi$, where $\psi$ represents the azimuth angle and $\chi$ denotes the ellipticity of the polarization state.

In the circular polarization basis, the unitary operator governing a lossless Hermitian optical system, represented by its Jones matrix, can be expressed in terms of the Pauli spin matrices as:

$$\hat{U} = |n_+\rangle\langle n_+|e^{i\phi_+} + |n_-\rangle\langle n_-|e^{i\phi_-} = \exp\left(i\frac{\delta}{2}\hat{n}\cdot\hat{\boldsymbol{\sigma}}^* + i\boldsymbol{\phi}\right)$$

$$= e^{i\phi}\begin{pmatrix} \cos\dfrac{\delta}{2} + i\sin\dfrac{\delta}{2}\sin 2\chi & i\sin\dfrac{\delta}{2}\cos 2\chi\, e^{i2\psi} \\ i\sin\dfrac{\delta}{2}\cos 2\chi\, e^{-i2\psi} & \cos\dfrac{\delta}{2} - i\cos\dfrac{\delta}{2}\sin 2\chi \end{pmatrix} = \begin{pmatrix} t_{++} & t_{+-} \\ t_{-+} & t_{--} \end{pmatrix} \tag{S3}$$

where the dynamical phase $\phi = \dfrac{\phi_+ + \phi_-}{2}$ and the retardance (birefringence of the system) $\delta = \phi_+ - \phi_-$, are the mean and difference eigenvalues' phase $\phi_+$ and $\phi_-$, and $\hat{\boldsymbol{\sigma}} = (\hat{\sigma}_x, \hat{\sigma}_y, \hat{\sigma}_z)$ is the vector of Pauli spin matrices consisting of three basic spin matrices. Please note that the application of sigma conjugation ensures a positive azimuthal rotation (anti-clockwise) of linear/elliptical polarization states, which is consistence with the phase factors considered in circular basis states in Eq. S2(a-b).

$$\hat{\sigma}_x = \begin{pmatrix} 0 & 1 \\ 1 & 0 \end{pmatrix}, \hat{\sigma}_y = \begin{pmatrix} 0 & -i \\ i & 0 \end{pmatrix}, \hat{\sigma}_z = \begin{pmatrix} 1 & 0 \\ 0 & -1 \end{pmatrix} \tag{S4}$$

When a normal incident wave of polarization $|\Psi_0\rangle$ interacts with an eigen system described by Eq. S3, the resulting output state is given by $|\Psi\rangle = \hat{U}|\Psi_0\rangle$. The co-polarized transmittance $(t_{co})$, is defined as the inner product of the input and output polarization states:

$$t_{co} = |t_{co}|e^{i\zeta} = \langle\Psi_0|\hat{U}|\Psi_0\rangle \tag{S5}$$

Let us assume the RHCP wave as input, i.e. $|\Psi_0\rangle = |R\rangle$. In this case, the complex transmittance in the co-polarized channel is $t_{co} = \cos\dfrac{\delta}{2} + i\sin\dfrac{\delta}{2}\sin 2\chi$, indicating that $t_{co}$ is governed by the eigen birefringence $\delta$ and eigen ellipticity $\chi$. Herein, the coordinate in eigen parameter space $\mathbf{R} = (\delta, \chi)$ characterizes the property of evolution operator $\hat{U}$, and thereby controlling the dynamics of complex transmission amplitude in co-polarized channels. Following a path in eigen parameter space, the accumulated phase along the path corresponds to a set of meta-atoms, which can be designed to engineer the phase response of the co-polarized wave. Therefore, it is necessary to characterize the phase difference between any points along the path. In other words, a connection needs to be constructed to compare the phase between arbitrary two points along the path. As the two points $\mathbf{R}$ and $\mathbf{R} + d\mathbf{R}$ in the parameter space are closely adjacent ($d\mathbf{R} \rightarrow 0$), the phase difference $(\Delta\zeta(\mathbf{R}))$ is

$$e^{i\Delta\zeta(\mathbf{R})} = \frac{e^{i\zeta(\mathbf{R}+d\mathbf{R})}}{e^{i\zeta(\mathbf{R})}} = \frac{\langle\Psi_0|\hat{U}(\mathbf{R}+d\mathbf{R})|\Psi_0\rangle}{\langle\Psi_0|\hat{U}(\mathbf{R})|\Psi_0\rangle} \times \frac{|\langle\Psi_0|\hat{U}(\mathbf{R})|\Psi_0\rangle|}{|\langle\Psi_0|\hat{U}(\mathbf{R}+d\mathbf{R})|\Psi_0\rangle|} \tag{S6}$$

According to equivalent infinitesimal, there is

$$\frac{\left|\langle\Psi_0\left|\widehat{U}(\mathbf{R})\right|\Psi_0\rangle\right|}{\left|\langle\Psi_0\left|\widehat{U}(\mathbf{R}+\mathrm{d}\mathbf{R})\right|\Psi_0\rangle\right|} \to 1, e^{ix} \to 1 + ix \tag{S7}$$

Eq. S7 turns to

$$e^{i\Delta\zeta(\mathbf{R})} \approx \frac{\langle\Psi_0\left|\widehat{U}(\mathbf{R}+\mathrm{d}\mathbf{R})\right|\Psi_0\rangle}{\langle\Psi_0\left|\widehat{U}(\mathbf{R})\right|\Psi_0\rangle} \to 1 + i\Delta\zeta(\mathbf{R}) \tag{S8}$$

The middle term in Eq. S8 can be expanded according to the Taylor series

$$\frac{\langle\Psi_0\left|\widehat{U}(\mathbf{R}+\mathrm{d}\mathbf{R})\right|\Psi_0\rangle}{\langle\Psi_0\left|\widehat{U}(\mathbf{R})\right|\Psi_0\rangle} = \frac{t_{co}(\mathbf{R}+\mathrm{d}\mathbf{R})}{t_{co}(\mathbf{R})} \to 1 + \frac{\frac{\partial t_{co}(\mathbf{R})}{\partial\mathbf{R}}\cdot\mathrm{d}\mathbf{R}}{t_{co}(\mathbf{R})} + \cdots \tag{S9}$$

Combining Eq. S8 and S9, $\Delta\zeta(\mathbf{R})$ can be derived as

$$\Delta\zeta(\mathbf{R}) = -i\frac{\langle\Psi_0\left|\frac{\partial\widehat{U}(\mathbf{R})}{\partial\mathbf{R}}\right|\Psi_0\rangle}{\langle\Psi_0\left|\widehat{U}(\mathbf{R})\right|\Psi_0\rangle}\cdot\mathrm{d}\mathbf{R} = Im\left(\frac{\partial}{\partial\boldsymbol{R}}\ln\langle\Psi_0\left|\widehat{U}(\mathbf{R})\right|\Psi_0\rangle\right)\cdot\mathrm{d}\mathbf{R} \tag{S10}$$

The phase difference between any two points along the path can be expressed as the path integral of the Berry connection term $\mathcal{A}_\mathbf{0}(\mathbf{R}) = i\frac{\partial}{\partial\mathbf{R}}\ln\langle\Psi_0|\widehat{U}(\mathbf{R})|\Psi_0\rangle$. Accordingly, the associated geometric Berry phase in the parameter space is defined as:

$$\phi_G(\boldsymbol{R}) = \int_{\mathbf{R_1}}^{\mathbf{R_2}}\mathcal{A}_\mathbf{0}(\mathbf{R})\cdot\mathrm{d}\mathbf{R} \tag{S11}$$

The accumulated geometric phase $\phi_G = -\zeta(\mathbf{R})$ (with the negative sign convention) depends on the curvature of the path taken from $\boldsymbol{R}_1$ to $\boldsymbol{R}_2$ in the eigen parameter space $\mathbf{R}$, which characterizes the instantaneous eigen states of the system. This formulation closely resembles the classical geometric Berry phase, where phase accumulation is derived from the instantaneous evolution of eigenstates in configuration space. However, a subtle difference in this approach is that instead of analyzing the evolution of the system's eigenstates directly, we track down the evolution of an input polarization governed by the unitary operator, expressed as $|\Psi\rangle = \widehat{U}(\mathbf{R})|\Psi_0\rangle$. And $|\Psi\rangle$ is subsequently projected onto the same initial input state $|\Psi_0\rangle$ using the projection operator $\widehat{P} = |\Psi_0\rangle\langle\Psi_0|$ on the Poincare sphere. Therefore, the accumulation of geometric phase in the co-polarization transmission channel depends on the eigen parameters $\mathbf{R} = (\delta, \chi)$, which can also be mapped at the Stokes parametric position of $|\Psi\rangle$ on the surface of the Poincare sphere, as illustrated in the next section.

## Supplementary Note 2: Manifestation of hidden singularity on the surface of the Poincaré sphere

Let us denote the output state $|\Psi(\mathbf{R}_s)\rangle = \widehat{U}(\mathbf{R})|\Psi_0\rangle$, where $\mathbf{R}_s$ represents the angular coordinates of $|\Psi\rangle$ on the Poincaré sphere. Eq. S5 has shown that the co-polarized transmittance $t_{co}$ only depends on eigen parameters: birefringence $\delta$ and ellipticity $\chi$ and is free from azimuth $\psi$. In other words, $|\Psi_0\rangle$ is acted by the operator after eliminating the influence of azimuth $\psi$ (azimuth $\psi$ can be regarded as a constant), and $|\Psi(\mathbf{R}_s)\rangle$ is the partial evolution state of the actual evolution state after removing the effect of azimuth angle. For simplicity, when $\psi$ is fixed at $-\pi/2$ to set $t_{-+}$ a real number, the output state $|\Psi(\mathbf{R}_s)\rangle$ can be mapped on the Poincaré sphere according to:

$$|\Psi(\mathbf{R}_s)\rangle = \left(\frac{t_{co}}{\sqrt{1-|t_{co}|^2}}\right) = \begin{pmatrix} \cos\dfrac{\delta}{2} + i\sin\dfrac{\delta}{2}\sin 2\chi \\ \sin\dfrac{\delta}{2}\cos 2\chi \end{pmatrix} = \begin{pmatrix} \cos\left(\dfrac{\pi}{4}-\chi_s\right)e^{i2\psi_s} \\ \sin\left(\dfrac{\pi}{4}-\chi_s\right) \end{pmatrix} \tag{S12}$$

The azimuth angle $2\psi_s$ and elevation angle $2\chi_s$ on the Poincaré sphere are:

$$2\chi_s = 2\left(\frac{\pi}{4} - \text{atan}\frac{\sqrt{1-|t_{co}|^2}}{|t_{co}|}\right) \tag{S13a}$$

$$2\psi_s = \arg(t_{co}) \tag{S13b}$$

where arg() denotes the argument function. The mapping between $(2\chi_s, 2\psi_s)$ and $(\delta, \chi)$ is depicted in Fig. S2A and B, where the angular coordinate in the parameter space can completely cover Poincaré sphere. So far, the complete mapping between $(2\chi_s, 2\psi_s)$ and $(\delta, \chi)$ is established. Next, the transmission amplitude and phase at each $(\delta, \chi)$ in parameter space are mapped onto the Poincaré sphere according to the mapping between $(2\chi_s, 2\psi_s)$ and $(\delta, \chi)$ in Fig. S2A and B.

Fig. S2C clearly shows that the singularity of amplitude in the parameter space is exactly at the south pole of Poincaré sphere, the polarization state $|L\rangle$, which is the antipole of the RHCP incident wave. The phase around the singularity changes by $2\pi$ continuously along the parallel but constant along the meridian. Consequently, the phase difference between any two closely adjacent points ($d\mathbf{R}_s \to 0$) on the sphere is

$$e^{i\Delta\zeta(\mathbf{R}_s)} = \frac{\langle\Psi_0|\Psi(\mathbf{R}_s+d\mathbf{R}_s)\rangle}{\langle\Psi_0|\Psi(\mathbf{R}_s)\rangle} \to 1 + i\Delta\zeta(\mathbf{R}_s) \tag{S14a}$$

$$\Delta\zeta(\mathbf{R}_s) = -i\frac{\left\langle\Psi_0\left|\dfrac{\partial}{\partial\mathbf{R}_s}\right|\Psi(\mathbf{R}_s)\right\rangle}{\langle\Psi_0|\Psi(\mathbf{R}_s)\rangle}\cdot d\mathbf{R}_s = \text{Im}\frac{\left\langle\Psi_0\left|\dfrac{\partial}{\partial\mathbf{R}_s}\right|\Psi(\mathbf{R}_s)\right\rangle}{\langle\Psi_0|\Psi(\mathbf{R}_s)\rangle}\cdot d\mathbf{R}_s \tag{S14b}$$

According to the phase distribution in Fig. S2D, the derivative of phase on the sphere is

$$\text{Im}\frac{\left\langle\Psi_0\left|\dfrac{\partial}{\partial\mathbf{R}_s}\right|\Psi(\mathbf{R}_s)\right\rangle}{\langle\Psi_0|\Psi(\mathbf{R}_s)\rangle} = \frac{1}{\cos 2\chi}\hat{e}_{2\psi} \tag{S15a}$$

$$d\mathbf{R}_s = \cos 2\chi\, d2\psi(\mathbf{R}_s)\hat{e}_{2\psi} + d2\chi(\mathbf{R}_s)\hat{e}_{2\chi} \tag{S15b}$$

where $\hat{e}_{2\psi}$ and $\hat{e}_{2\chi}$ are the orthogonal unit vectors along azimuthal and elevation direction, $2\psi(\mathbf{R}_s)$ and $2\chi(\mathbf{R}_s)$ are the azimuthal and elevation angles of point $\mathbf{R}_s$.

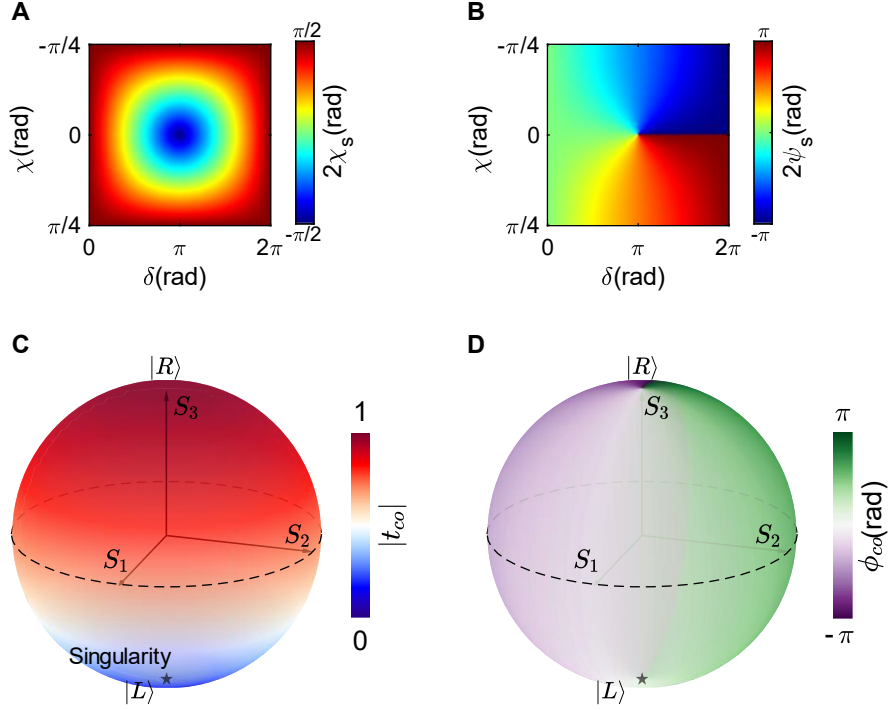

**Fig. S2. Map complex amplitude $t_{co}$ from parameter space to Poincaré sphere.** (**A**) The elevation angle $2\chi_s$ and (**B**) azimuth angle $2\psi_s$ in the eigen parameter space of $(\delta, \chi)$. (**C**) The amplitude and (**D**) phase of $t_{co}$ mapped onto Poincaré sphere, the singularity (star mark) is on the south pole and the maximum amplitude is on the north pole, the singularity is always at the antipole of the maximum amplitude. The phase around the singularity shifts by $2\pi$ along parallel continuously and remains unchanged along the meridian.

Consequently, by substituting Eq. S15 into Eq. S14b and then integrate $\Delta\zeta(\mathbf{R}_s)$ along a path starting from an initial point $\mathbf{R}_0$ to target point $\mathbf{R}_s$ the integration can be simplified as

$$\zeta(\mathbf{R}_s) = \int_{\mathbf{R}_0}^{\mathbf{R}_s} \Delta\zeta(\mathbf{R}_s) \cdot d\mathbf{R} = \int_{\mathbf{R}_0}^{\mathbf{R}_s} 1 d2\psi(\mathbf{R}_s) = 2\psi(\mathbf{R}_s) - 2\psi(\mathbf{R}_0) \tag{S16}$$

Eq. S16 shows that the integration only depends on the azimuthal angle of the starting and ending points, which is purely geometric. To demonstrate the geometric property of the phase accumulation, we calculate the phase accumulation along three different smooth closed loops outside (solid circle), crossing (dashed circle) and encircling (dotted circle) the singularity, which are similar to the curves in Fig. 2 in the main text (Fig. S3A and B). For the sake of clarity, here we denote the angle of the circle on parameter space as $\alpha$, and the paths on Poincaré sphere are projected to the equatorial plane (Fig. S3C) because the phase accumulation along the paths on the sphere only depends on the azimuthal angle.

The phase accumulation result in Fig. S3D is directly calculated by Eq. S5 and acts as a reference. Figs. S3E and F show the phase results calculated by the integration of Eqs. S14b and 16, respectively. Obviously, the phase curves in Fig. S3E and F are the same as the reference results in Fig. S3D, where the phase accumulation range of the solid, dashed and dotted curves is less than $\pi$, equal to $\pi$ and $2\pi$, respectively. The results validate Eq. S11 and Eq. S16 and demonstrate that encircling the singularity can gain $2\pi$ phase accumulation, indicating the topologically protected property of the co-polarized singular phase.

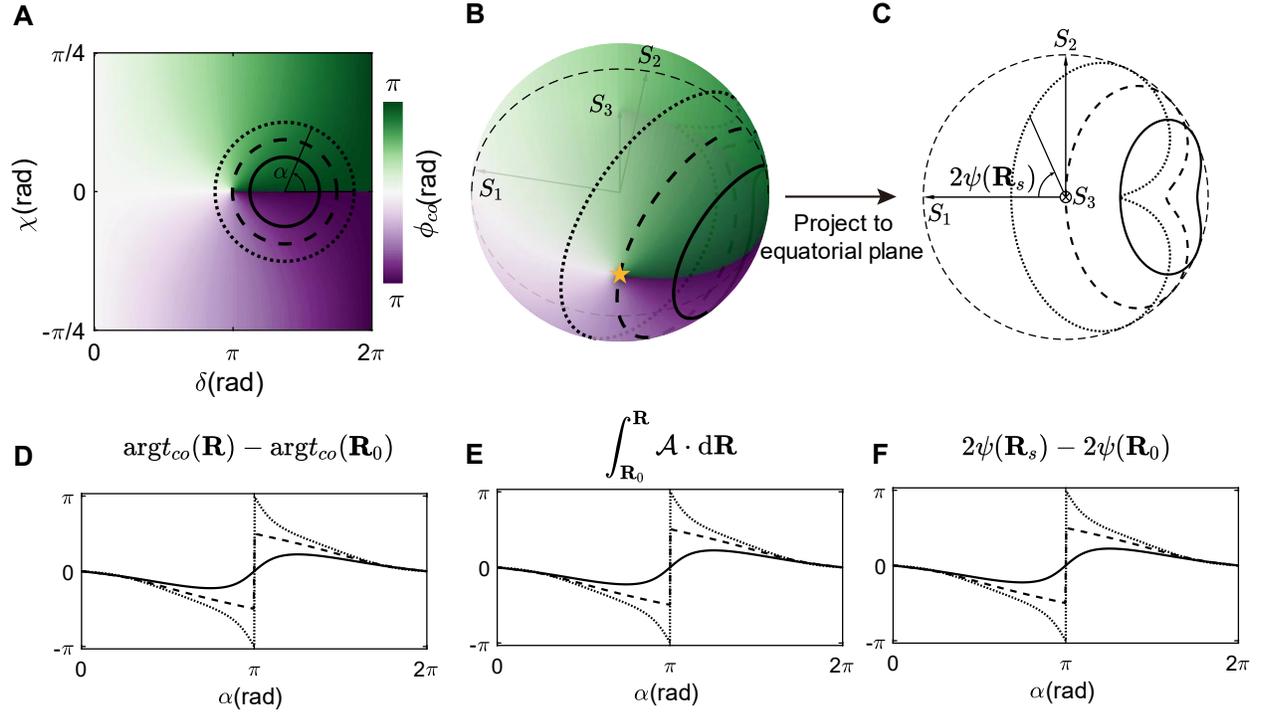

**Fig. S3. Phase accumulation along loops outside, crossing and encircling the singularity.** Three different smooth closed loops outside (solid circle), crossing (dashed circle) and encircling (dotted circle) the singularity in (**A**) parameter space and (**B**) Poincaré sphere. (**C**) The projected paths on the equatorial plane. The phase accumulation along the three paths calculated by (**D**) Eq. S5, (**E**) Eq. S11 and (**F**) Eq. S16.

**Supplementary Note 3: Co-polarized singular phase and associated topology protection**

To understand the behaviour of the co-polarized singular phase and its associated topology protection, we first plot the co-polarized phase in the eigen parameter space $\mathbf{R} = (\delta, \chi)$ as we have discussed in the main text (Fig. S4A) in great details. This phase is then mapped onto a toroidal surface $T^2$ to capture its periodicity and topological properties. In Fig. S4B, the toroidal surface $T^2$ is visualized by considering $\delta$ as the azimuthal angle and $\phi_{co}$ as the poloidal angle of the torus surface, respectively. The dashed loop represents a closed trajectory in the eigen parameter space that does not enclose the singularity, leading to a total phase accumulation along the loop is zero. Since the loop can be smoothly deformed out of the torus, it corresponds to a trivial topology with a Chern number of $C = 0$, indicating no topological protection.

In contrast, the solid loop encloses the singularity and accumulates a total phase of $-2\pi$ as it encircles the singular point in an anticlockwise direction. This accumulation is a hallmark of a nontrivial topology, as the loop cannot be smoothly deformed out of the torus without cutting through it. Consequently, it exhibits a topologically protected phase with a Chern number of $C = -1$, ensuring a robust and complete $2\pi$ phase accumulation in the co-polarized transmission channel.

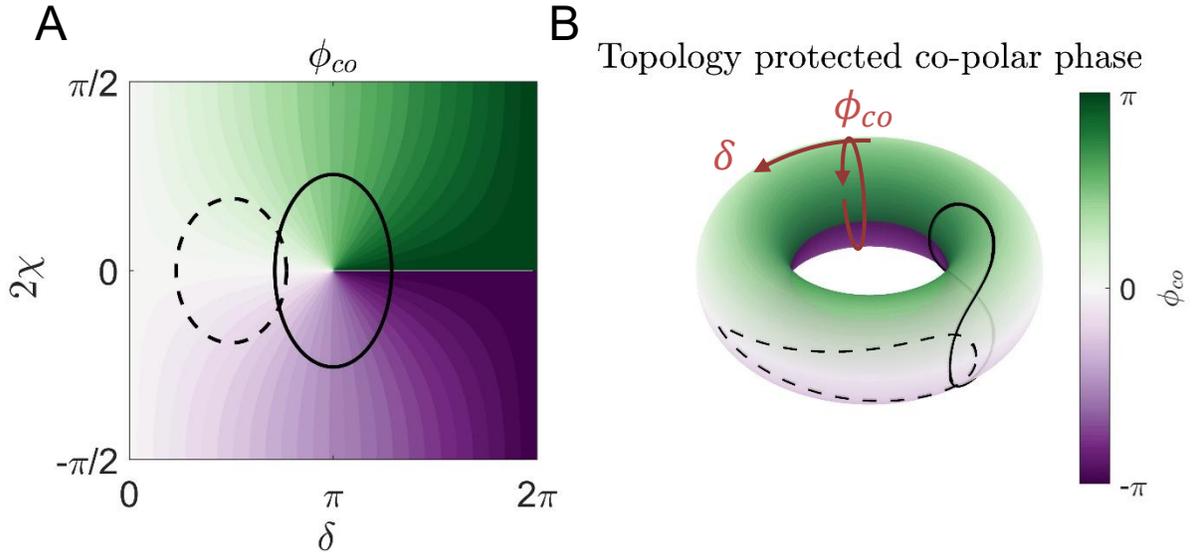

**Fig. S4. The enclosing singularity and associated topological protection.** (**A**) Two closed loops outside (dashed) and encircling (solid) the co-polarized phase singularity are shown in the eigen parameter space. (**B**) Topological depiction of the co-polarized phase on a toroidal surface. The phase variation along the solid loop is constrained by the toroidal topology, with a Chern number of $C = -1$, leading to a full $2\pi$ phase accumulation in the co-polar transmission channel.

## Supplementary Note 4: Encircle the singularity of the co-circular channels with maximum efficiency by circular birefringent meta-atom

It has been demonstrated that encircling the singularity enables achieving $2\pi$ phase accumulation. It is also necessary to take into account the amplitude along the path since it determines the transmission efficiency. Fig. S5A shows the amplitude on the Poincaré sphere clearly, the larger the elevation angle the higher the amplitude because $|t_{co}| = \cos(\pi/4 - \chi_s)$. For instance, when the evolution loop is extremely close to the north pole, the amplitude on this path is almost unity, but the length of the curve is extremely short (the very tiny green circle near the north pole) making it look like a point and is hardly observable on the sphere. To clearly show the amplitude and phase result simultaneously, the sphere is expanded according to the relationship between the angular coordinate and complex amplitude as follows:

$$\begin{cases} x = |t_{co}| \cos 2\psi_s \\ y = |t_{co}| \sin 2\psi_s \\ z = |t_{co}| = \cos\left(\dfrac{\pi}{4} - \chi_s\right) \end{cases} \tag{S17}$$

The expanded sphere is shown in Fig. S5B, where the conical surface represents the amplitude and the corresponding phase is projected on the heatmap. The amplitude singularity at the south pole of the sphere turns to the tip of the conical surface, while the phase around the singularity shows a continuous variation of $2\pi$. The extremely short evolution loop on the sphere corresponds to the top edge of the conical surface, which holds a unity amplitude as well as linear $2\pi$ phase accumulation, i.e. $t_{++} = \pm e^{i\delta/2}$, and the phase shift is linear to $\delta/2$. According to Eq. S5, there should be $\chi = \pm\dfrac{\pi}{4}$ and $\delta \in [0, 2\pi]$, indicating that the eigenstate of the meta-atom should agree with the incident polarization and the corresponding eigen birefringence needs to be tuned within 0 to $2\pi$.

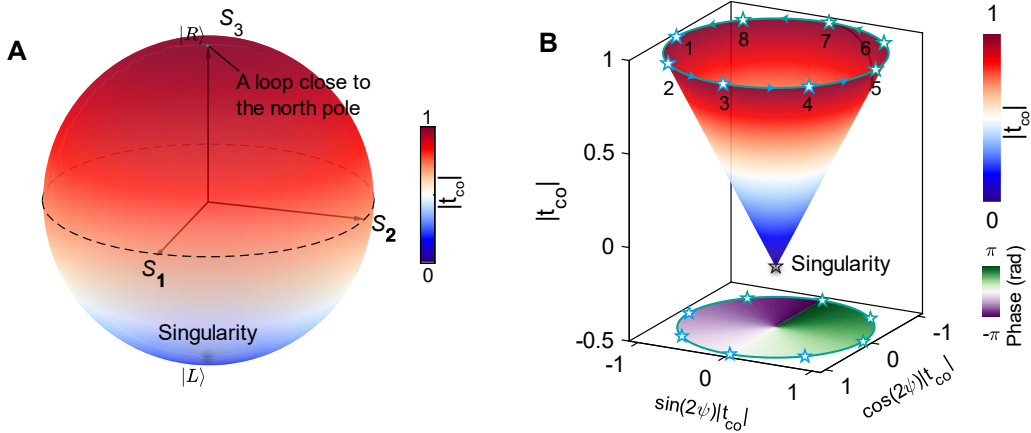

**Fig. S5. The enclosing path with amplitude near unity on the Poincaré sphere and the expanded Poincaré sphere.** (**A**) The loop path with amplitude approaching unity is extremely close to the north pole, which is shown in an extremely short green loop that is hard to observe. (**B**) Amplitude (conical surface) and phase (bottom heat map) of $t_{co}$ on the expanded Poincaré sphere, the loop path with maximum amplitude is the top edge of the conical surface, and the phase along the path shifts by $2\pi$ continuously. The numbered discrete points correspond to the element in Fig. S6B-E.

To encircle the singularity around the top edge of the conical surface under CP incident wave, we design a circular birefringent meta-atom exhibiting high transmission. The detailed schematic structure of the proposed meta-atom is shown in Fig. S6A, which consists of two 2.5 mm thick FB430 dielectric substrates ($\varepsilon_r$ = 4.3, tan $\delta$ = 0.002), a 0.1 mm thick bonding layer ($\varepsilon_r$ = 3.5, tan $\delta$ = 0.004), three layers of printed copper pattern and four vertical metal vias. Both the top and bottom copper layers contain four identical patches forming $C_{4z}$ symmetry, which are electrically connected by metal vias to build four patch-via-patch structures that allow direct coupling. Meanwhile, the middle copper layer is a metal ground plane with circular trims (the radius is 0.15 mm larger than the vias) allowing vias to penetrate without any contact with the ground plane. The ground plane allows the patches facing the incidence to directly couple with the incident wave while shielding the patch on another side. As a result, when the EM wave passes through the metasurface, the patch layer facing the incidence acts as a receiver, and the received EM energy is coupled to the patch layer on the other side by the vias, which acts as a transmitter to radiate the energy to free space on the transmission side. Relatively rotating the top and bottom patches connected by the same via can be effectively regarded as changing the polarization angle of the antennas, which causes a circular birefringence $\delta$ nearly twice the counterclockwise rotation angle $\theta$ ($\delta \approx 2\theta$), and the co-polarized singular phase of the RHCP and LHCP transmitted wave is $\delta/2$ and $-\delta/2$ respectively.

The goal is to engineer the structure to gain $2\pi$ accumulation of co-polarized singular phase. For the designed structure, the neighbouring patches may touch each other when the rotation angle $\theta$ is over 90°, which will sharply change the property of the meta-atom and deteriorate the performance. To avoid this issue, the bottom patches are fixed and the top patches are rotated counterclockwise around their metal vias within 0°-90°, which is sufficient to tune $\delta$ from 0 to $\pi$ (quarter circle from elements 3 to 5 in Fig. S5) and allows phase modulation of both RHCP and LHCP to realize $\phi_{++} = -\phi_{--} \in [0, \pi/2]$, as shown in Fig. S6C. Subsequently, mirroring the element with respective to the $xoz$ plane (or $yoz$ plane) effectively inverts the sign of $\chi$ while maintaining $\delta$ unchanged. Further combined with $\delta$ tuning from 0 to $\pi$, phase coverage within [-$\pi/2$, 0] is achieved (Fig. S6B), thereby extending the total phase range to [-$\pi/2$, $\pi/2$], complete half-circle loop in the phase diagram in Fig. S5 from elements 1 to 5.

Next, we need to fill the phase range within [$\pi/2$, $\pi$] and [-$\pi$, -$\pi/2$] to cover the full phase range. The phase range within [$\pi/2$, $\pi$] is equal to adding $\pi$ to the interval [-$\pi/2$, 0], hence, denotes the co-polarized singular phase within the two different ranges as $\phi_{++} = -\phi_{--} \in [-\pi/2, 0]$ and $\phi'_{++} = -\phi'_{--} \in [\pi/2, \pi]$, they follow the relationship as below:

$$\phi'_{++} = \phi_{++} + \pi, \phi'_{--} = \phi_{--} - \pi \tag{S18}$$

The corresponding circular birefringence and mean phase are

$$\delta' = \delta + 2\pi, \phi' = \phi = 0 \tag{S19}$$

Substituting them into the operator in Eq. S3, there is

$$\widehat{U}(\delta') = \exp\left(i\frac{\delta + 2\pi}{2}\hat{n} \cdot \widehat{\boldsymbol{\sigma}}^*\right) = -\widehat{U}(\delta) \tag{S20}$$

The above equation indicates that there is a polarization independent $\pi$-phase difference between the elements with phase covering in the [-$\pi/2$, 0] range and those in the [$\pi/2$, $\pi$] range, which is equivalent to adding a π dynamic phase. Therefore, the distinct geometric size of elements in Fig. S6B and D aims to add a global π-phase difference, and the same applies to elements in Fig. S6C and E. By tuning the elements in Fig. S6D and E in the same way, the phase range within [π/2, π] and [-π, -π/2] is achieved respectively. So far, the engineered elements can completely encircle the singularity to gain 2π accumulation.

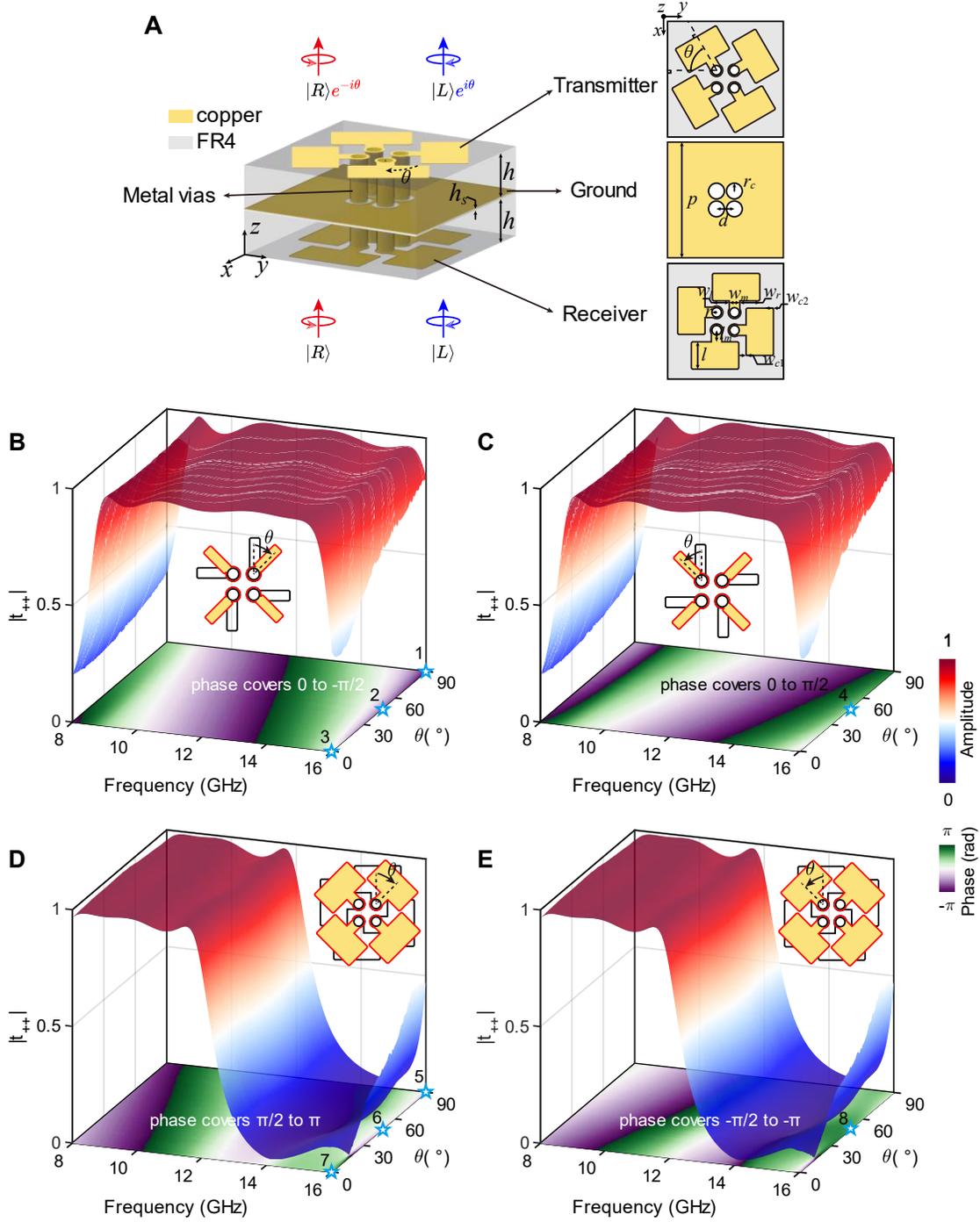

**Fig. S6. Tunning circular birefringent meta-atom to completely cover 2π phase accumulation.** (**A**) The structure and propagation schematic of meta-atom. The meta-atom consists of four parts: receiver patch, transmitter patch, metal vias, and ground plane. The patches on the receiver and transmitter layer are electrically connected by a via, and the relative rotation

angle $\theta$ between them can be tuned from 0° to 90°. As the RHCP and LHCP wave pass through the metasurface, the patch layer facing the incidence acts as a receiver antenna, and the received EM energy is coupled to the patch layer on another side by the vias, which acts as a transmitter to radiate the energy to free space on the transmission side. The transmitted RHCP and LHCP gain a $-\theta$ and $\theta$ phase respectively (the schematic here is clockwise rotation, while the phase sign is inverted for the case of counterclockwise rotation. ). (**B**) The complex transmittance of meta-atom (inset shows its shape) covering the phase shift will decrease from 0 to $-\pi/2$ as the relative rotation angle $\theta$ increases from 0° to 90° clockwise, covering phase range of [$-\pi/2,0$] and obtaining elements 1 to 3. Mirror the structure with respective to *xoz* plane and increase $\theta$ from 0 to $\pi/2$ counterclockwise, (**C**) the amplitude is almost unchanged while the corresponding phase coverage becomes [0,$\pi/2$], and element 4 is realized. (**D**) By increasing the patch size to introduce a polarization-independent $\pi$ phase shift to elements 1 to 3, the spectrum undergoes a redshift and results in a corresponding phase range of [$\pi/2$, $\pi$] to obtain elements 5 to 7. (**E**) In the same way, mirror the structure in D to cover phase shift within [$-\pi,-\pi/2$] and obtain element 8.

Note that the transmission amplitude in Fig. S6B to E cannot perfectly reach unity, which is due to reflection rather than loss. The total reflection amplitude results corresponding to the transmission in Fig. S6B and D are shown in Fig. S7A and B. It is evident that the decrease of the transmission exactly corresponds to the increase of reflection, proving that the meta-atom is almost lossless and reflection reduces the structure's overall transmission. Furthermore, the medium property of the designed meta-atom is analyzed by generalized sheet transition conditions (GSTCs) theory (*28*) to illustrate its working mechanism.

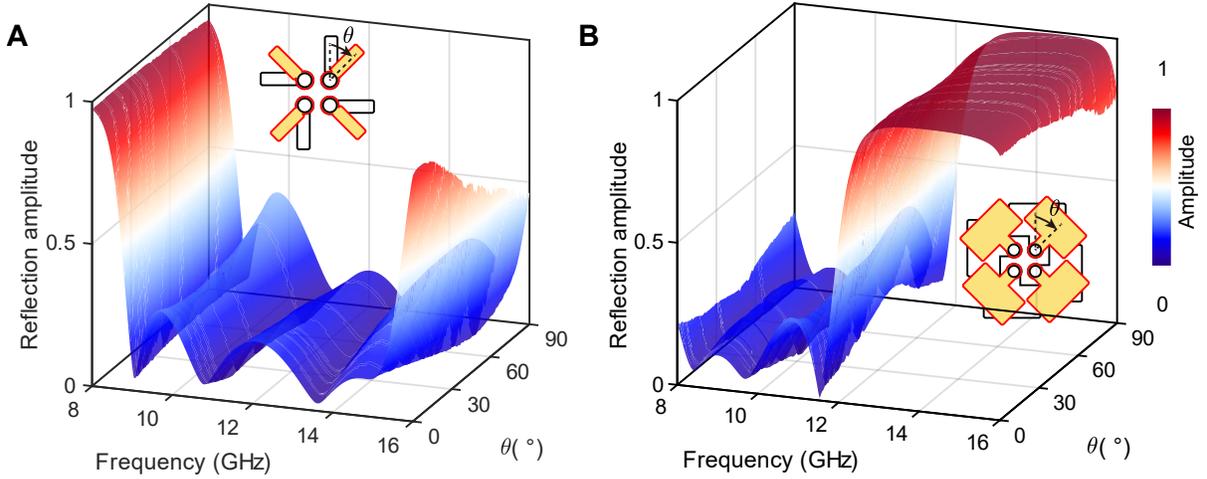

**Fig. S7. Reflection amplitude of the circular birefringent meta-atoms.** (A) The reflection amplitude of (**A**) the meta-atom in Fig.S6B and (**B**) Fig.S6D as the relative rotation angle $\theta$ is increased from 0 to $\pi/2$ clockwise.

According to GSTCs theory, when normal incident waves illuminate a two-dimensional (2D) uniform metasurface, the equivalent surface electric current and magnetic current ($\mathbf{J}^s$ and $\mathbf{M}^s$) can be induced on the metasurface. The relationship between the excitation field and the constitutive property of the meta-atom is described as:

$$\begin{pmatrix} \mathbf{J}^s \\ \mathbf{M}^s \end{pmatrix} = \begin{pmatrix} \boldsymbol{\chi}_{ee} & \boldsymbol{\chi}_{em} \\ \boldsymbol{\chi}_{me} & \boldsymbol{\chi}_{mm} \end{pmatrix} \begin{pmatrix} \mathbf{E}^{av} \\ \mathbf{H}^{av} \end{pmatrix} \tag{S21}$$

where $\mathbf{J}^s = [J_x^s \quad J_y^s]^T$ and $\mathbf{M}^s = [M_x^s \quad M_y^s]^T$, and $\mathbf{E}^{av} = [E_x^{av} \quad E_y^{av}]^T$, $\mathbf{H}^{av} = [H_x^{av} \quad H_y^{av}]^T$ are the average tangential electric and magnetic vector fields to surface, and superscript T denotes transpose. The $\boldsymbol{\chi}_{ee}$, $\boldsymbol{\chi}_{mm}$, $\boldsymbol{\chi}_{em}$, $\boldsymbol{\chi}_{me}$ are all the $2 \times 2$ tensors of electric, magnetic, magnetic-to-electric and electric-to-magnetic equivalent susceptibility, presenting the following form:

$$\boldsymbol{\chi}_{ee} = \begin{pmatrix} \chi_{ee}^{xx} & \chi_{ee}^{xy} \\ \chi_{ee}^{yx} & \chi_{ee}^{yy} \end{pmatrix}, \quad \boldsymbol{\chi}_{em} = \begin{pmatrix} \chi_{em}^{xx} & \chi_{em}^{xy} \\ \chi_{em}^{yx} & \chi_{em}^{yy} \end{pmatrix}$$
$$\boldsymbol{\chi}_{me} = \begin{pmatrix} \chi_{me}^{xx} & \chi_{me}^{xy} \\ \chi_{me}^{yx} & \chi_{me}^{yy} \end{pmatrix}, \quad \boldsymbol{\chi}_{mm} = \begin{pmatrix} \chi_{mm}^{xx} & \chi_{mm}^{xy} \\ \chi_{mm}^{yx} & \chi_{mm}^{yy} \end{pmatrix}$$

(S22)

The four tensors totally contain 16 equivalent surface susceptibilities that can characterize the 2D constitutive property of the meta-atom. When the meta-atom is lossless, $\boldsymbol{\chi}_{ee}$ and $\boldsymbol{\chi}_{mm}$ are purely imaginary while $\boldsymbol{\chi}_{em}$ and $\boldsymbol{\chi}_{me}$ are purely real. Therefore, we will focus on the imaginary part of $\boldsymbol{\chi}_{ee}$, $\boldsymbol{\chi}_{mm}$ and the real part of $\boldsymbol{\chi}_{em}$, $\boldsymbol{\chi}_{me}$ because the meta-atom is almost lossless. The 16 susceptibilities can be retrieved from the corresponding scattering parameters ($S$ parameters) of the meta-atom:

$$\mathbf{S}_{nm} = \begin{pmatrix} S_{nm}^{xx} & S_{nm}^{xy} \\ S_{nm}^{yx} & S_{nm}^{yy} \end{pmatrix}$$

(S23)

where subscripts $n$ and $m = 1$, 2, denoting the propagation direction $m \rightarrow n$ (Fig. S8), and the superscripts $xx$ and $yx$ ($yy$ and $xy$) denote co- and cross-polarized responses under $x$- ($y$-) linear polarization (LP) normal incidence respectively.

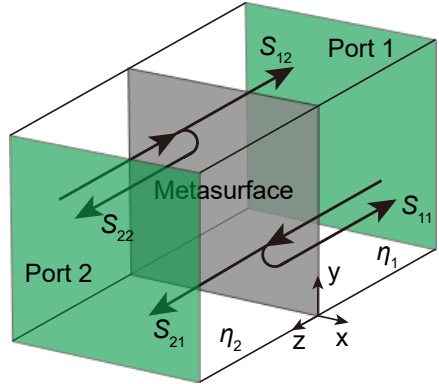

**Fig. S8. Normal propagation schematic of the general lossless metasurface**. Port 1 and port 2 are located on the two sides of the metasurface. $\eta_1$ and $\eta_2$ represent the wave impedance of the two regions respectively, and the arrows denote the propagation direction of the wave.

Then, the susceptibilities can be calculated from $S$ parameters according to (*29*)

$$\begin{pmatrix} \boldsymbol{\chi}_{ee} & \boldsymbol{\chi}_{em} \\ \boldsymbol{\chi}_{me} & \boldsymbol{\chi}_{mm} \end{pmatrix} = 2 \begin{pmatrix} \dfrac{\mathbf{I}}{\eta_1} - \dfrac{\mathbf{S}_{11}}{\eta_1} - \dfrac{\mathbf{S}_{21}}{\eta_2} & \dfrac{\mathbf{I}}{\eta_2} - \dfrac{\mathbf{S}_{12}}{\eta_1} - \dfrac{\mathbf{S}_{22}}{\eta_2} \\ \mathbf{N} + \mathbf{NS}_{11} - \mathbf{NS}_{21} & -\mathbf{N} + \mathbf{NS}_{12} - \mathbf{NS}_{22} \end{pmatrix} \times$$

$$\begin{pmatrix} \mathbf{I} + \mathbf{S}_{11} + \mathbf{S}_{21} & \mathbf{I} + \mathbf{S}_{12} + \mathbf{S}_{22} \\ \dfrac{\mathbf{N}}{\eta_1} - \dfrac{\mathbf{NS}_{11}}{\eta_1} + \dfrac{\mathbf{NS}_{21}}{\eta_2} & -\dfrac{\mathbf{N}}{\eta_2} - \dfrac{\mathbf{NS}_{12}}{\eta_1} + \dfrac{\mathbf{NS}_{22}}{\eta_2} \end{pmatrix} \tag{S24}$$

where $\eta_1$ and $\eta_2$ are the wave impedance in regions 1 and 2, respectively, $\mathbf{I} = \begin{pmatrix} 1 & 0 \\ 0 & 1 \end{pmatrix}$ and $\mathbf{N} = \begin{pmatrix} 0 & -1 \\ 1 & 0 \end{pmatrix}$ are the identity matrix and 90° rotation matrix, respectively. Since the lossless metasurface is placed in free space, both $\eta_1$ and $\eta_2$ are equal to the wave impedance of free space $\eta_0$. It must be noted that the $S$ parameters in Eq. S24 is under LP basis, and hence, the ideal Jones matrix of the circular birefringent meta-atom $\widehat{U} = \begin{pmatrix} e^{\frac{i\delta}{2}} & 0 \\ 0 & e^{-\frac{i\delta}{2}} \end{pmatrix} e^{i\phi}$ needs to be converted from CP basis to LP basis as follows (*30*):

$$\mathbf{S}_{21} = \widehat{\Lambda}\widehat{U}\widehat{\Lambda}^{-1} = \begin{pmatrix} \cos\dfrac{\delta}{2} & -\sin\dfrac{\delta}{2} \\ \sin\dfrac{\delta}{2} & \cos\dfrac{\delta}{2} \end{pmatrix} e^{i\phi} \tag{S25}$$

where $\widehat{\Lambda} = \begin{pmatrix} 1 & 1 \\ -i & i \end{pmatrix}$ is the basis transformation matrix. According to reciprocity and conservation of energy, there are $\mathbf{S}_{21} = \mathbf{S}_{12}^{\mathrm{T}}$ and $\mathbf{S}_{11} = \mathbf{S}_{22} = \mathbf{0}$. The mathematical from of Eq. S25 is the same as that of the rotation matrix $\begin{pmatrix} \cos\alpha & -\sin\alpha \\ \sin\alpha & \cos\alpha \end{pmatrix}$, indicating that the circular birefringent meta-atom acts as a polarization rotator to induce a counterclockwise rotation of the incident wave's polarization plane by $\delta/2$. For simplicity, replace $\delta/2$ in Eq. S25 by $\alpha$ and substitute all the $S$ parameters into Eq. S24, the ideal susceptibilities are obtained:

$$\boldsymbol{\chi}_{ee} = -\frac{i}{\eta_0}\frac{2\sin\phi}{\cos\alpha + \cos\phi}\mathbf{I} = \chi_{ee}\mathbf{I}, \qquad \boldsymbol{\chi}_{em} = \frac{2\sin\alpha}{\cos\alpha + \cos\phi}\mathbf{I} = \chi_{em}\mathbf{I}$$

$$\boldsymbol{\chi}_{me} = -\frac{2\sin\alpha}{\cos\alpha + \cos\phi}\mathbf{I} = \chi_{me}\mathbf{I}, \quad \boldsymbol{\chi}_{mm} = -i\eta_0\frac{2\sin\phi}{\cos\alpha + \cos\phi}\mathbf{I} = \chi_{mm}\mathbf{I} \tag{S26}$$

All the tensors present an identity matrix form, hence, and only 4 of the 16 susceptibilities in Eq. S22 are unique, which are denoted as $\chi_{ee}$, $\chi_{em}$, $\chi_{me}$, and $\chi_{mm}$, respectively. Due to the presence of magnetoelectric coupling and all the susceptibility tensors exhibiting identity matrix form, such medium is classically termed as a bi-isotropic medium (*31*).

Then, the relationship between $S$ parameters and the four unique susceptibilities is further analyzed, which can reveal the underlying relationship between constitutive and propagation characteristics. The $S$ parameters of the bi-isotropic meta-atom can be calculated reversely from susceptibilities as follows (*29*)

$$\begin{pmatrix} S_{11}^{xx} \\ S_{11}^{yx} \\ S_{21}^{xx} \\ S_{21}^{yx} \end{pmatrix} = C \begin{pmatrix} 2(\chi_{mm} - \eta_0^2 \chi_{ee}) \\ 2\eta_0(\chi_{me} + \chi_{em}) \\ (\chi_{me}\chi_{em} - \chi_{ee}\chi_{mm} + 4) \\ 2\eta_0(\chi_{me} - \chi_{em}) \end{pmatrix} \tag{S27}$$

where $C = \frac{1}{(\eta_0 \chi_{ee} + 2)(2\eta_0 + \chi_{mm}) - \eta_0 \chi_{me}\chi_{em}}$. According to Eq. S27, the linearly co- and cross-polarized reflection can be completely suppressed by engineering $\eta_0 \chi_{ee} = \chi_{mm}/\eta_0$ and $\chi_{em} = -\chi_{me}$. As for the linearly co- and cross-polarized, it is necessary to introduce and tune the magnetoelectric response delicately. When all the tensors are designed to agree with Eq. S26, we can achieve the reflectionless circular birefringent meta-atom perfectly.

Next, we characterize the designed meta-atom by GSTCs theory to demonstrate its excellent circular birefringence. Take meta-atom No. 1 in Fig. S6 (the same as No. 1 in the main text) as example, whose relative rotation angle $\theta = 90°$ and the ideal linearly cross-polarized transmittance reaches unity. Under the y-LP incident wave, the amplitude and phase of transmittance $S_{21}^{yy}$, $S_{21}^{xy}$ and reflectance $S_{11}^{yy}$, $S_{11}^{xy}$ are depicted in Fig. S9A and B, where the linearly cross-polarized transmittance $S_{21}^{xy}$ (red solid curve) is the dominant response across the operating bandwidth (8.8 GHz - 14.9 GHz), exhibiting an amplitude above 0.9 and accompanied by a smooth phase variation. Substituting the complex transmittance and reflectance into Eq. S24, the susceptibilities are obtained and shown in Fig. S9C and D, where the solid black curve is the susceptibilities under ideal conditions ($|S_{21}^{xy}| = 1$), providing a clear visual reference for evaluating the deviation of the realized susceptibilities from their ideal values.

Fig. S9C clearly shows that Im($\chi_{ee}\eta_0$) and Im($\chi_{mm}/\eta_{0e}$) maintain near-equality across a broad frequency band from 8.8 GHz to 15.9 GHz, which effectively suppresses the linearly co-polarized reflection and thus $|S_{11}^{yy}|$ is below 0.4. Meanwhile, the Re($\chi_{em}$) and Re($\chi_{me}$) in Fig. S9D are always opposite, leading to linearly cross-polarized reflectance $S_{11}^{xy}$ almost zero across the broad frequency band. When comparing the realized values with the ideal values (black curves), all the realized values are approximately equal to the ideal value in the operating bandwidth from 8.8 GHz to 14.9 GHz. As a result, both reflection and linearly co-polarized transmission are effectively suppressed in the specified frequency range, enabling the cross-polarized response to become the dominant response.

Through the GSTCs model, we have figured out that the equivalent surface susceptibilities of the designed meta-atom are close to the ideal ones, enabling excellent circular birefringence properties. In addition, we can qualitatively explain the relationship between the structure of meta-atom and resonance. (1) Structure's $C_{4z}$ symmetry (chiral four-fold rotation about the z-axis) ensures the isotropic constitutive property, which constrains the tensors in Eq. S22 to exhibit an identity form. (2) The vertical vias establish direct coupling between the connected patches, enabling the introduction of magnetoelectric coupling and results in non-zero values for both $\chi_{em}$ and $\chi_{me}$. (3) The metal ground plane not only effectively isolates the receiving and transmitting layers to enhance the robustness of the manipulation, but also supports current that is opposite to the patches to strengthen resonance, and consequently broaden the transmission bandwidth and improve the efficiency.

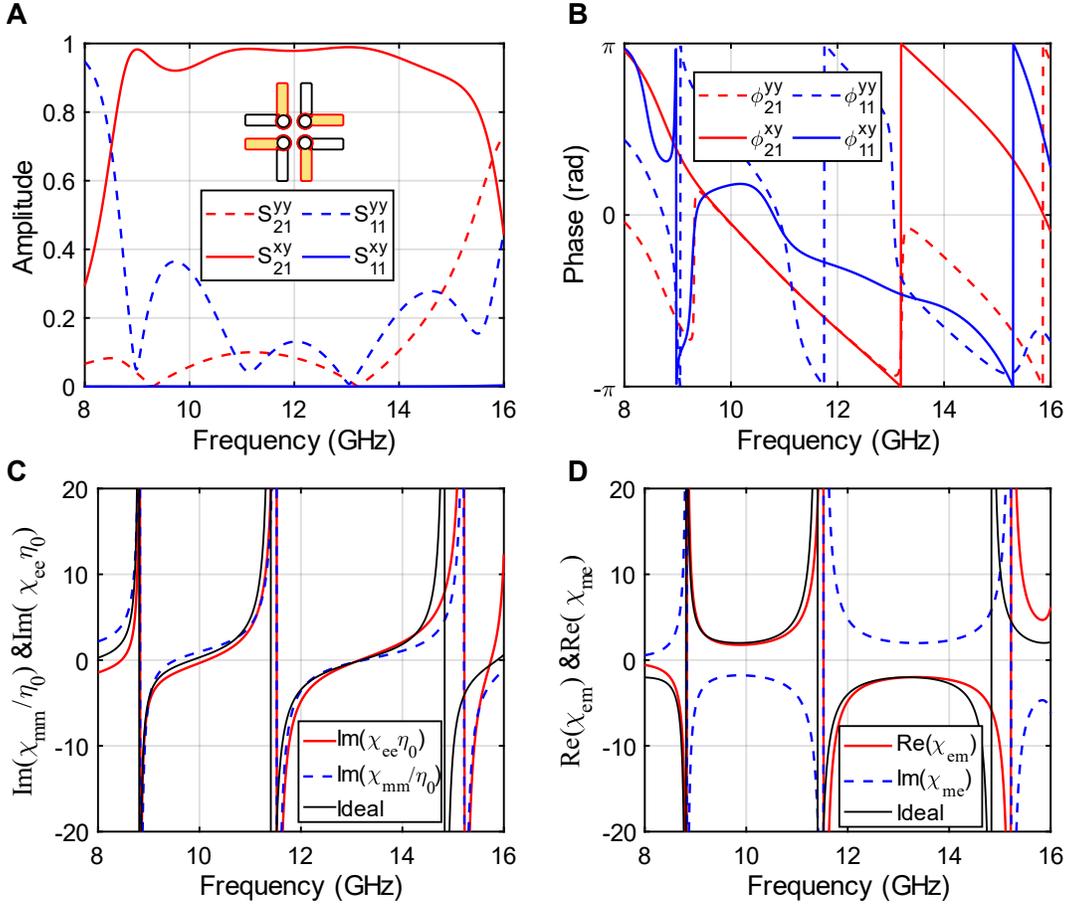

**Fig. S9. The *S* parameter and the Susceptibilities of the demonstration meta-atom.** (**A**) The amplitude and (**B**) phase of transmittance $S_{21}^{yy}$ (red dashed curve), $S_{21}^{xy}$ (red solid curve) and reflectance $S_{11}^{yy}$ (blue dashed curve), $S_{11}^{xy}$ (blue solid curve). The (**C**) imaginary part of $\chi_{ee}\eta_0$ (red solid curve) and $\chi_{mm}/\eta_0$ (blue dashed curve), (**D**) real part of $\chi_{em}$ (red solid curve) and $\chi_{em}$ (blue dashed curve), and the black solid curves are the ideal result when $S_{11}^{xy} = 1$.

**Supplementary Note 5: Fabrication of metasurface prototypes, measurement methods and transmission efficiency characterization**

Fabrication of metasurface prototypes

All the proposed metasurface prototypes are fabricated by printed circuit board (PCB) technology and consist of two 2.5 mm thick FB430 dielectric substrates. The copper films on each substrate are etched into the design patterns and then a 0.1 mm thick bonding layer is used to stick the dielectric substrates tightly. To form the vertical metal vias, holes are drilled in the PCBs and a thin layer of copper coating is deposited chemically on the inner wall of the holes.

Measurement methods

The experimental setup is schematically depicted in Fig. S10A. The receiving and transmitting antennas are both linearly polarized. To obtain the results under circular polarization states, both the co- and cross-polarized transmission fields under $x$- and $y$-LP incident waves are measured. Then the transmitted electric field for any polarization state can be obtained according to the polarization synthesis.

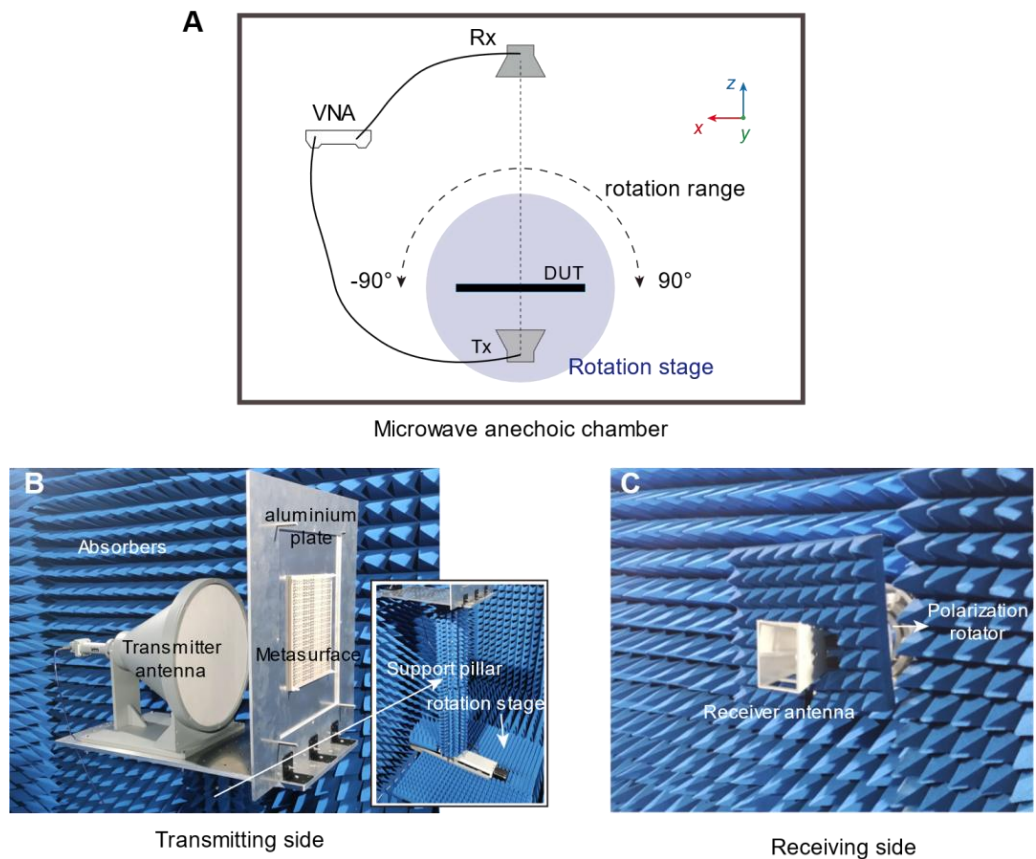

**Fig. S10. Experimental setup for metasurface measurement.** (A) The measurements are conducted in a microwave anechoic chamber, where the metasurface, transmitting (Tx) antenna and receiving (Rx) antenna are horizontally and vertically aligned. The antennas are connected to a Keysight P9384B vector network analyzer (VNA) to extract the scattering parameters. (B) On the transmission side, a rotation stage supporting a turn table to hold the metasurface and transmitter antenna is used. The metasurface under test is placed at the rotation center of the stage and the dielectric lens-loaded high gain horn antenna placed 200 mm in front of the metasurface

is used as the illumination source. On the receiving side, a dual-ridged horn antenna placed 2.5 m (83λ at 10 GHz) away from the metasurface is used to collect the transmitted intensities.

<u>Transmission efficiency characterization</u>

Based on the measured far-field results of the two metasurface prototypes, their transmission efficiencies are calculated to characterize the high efficiency merit:

$$T_{ij} = \frac{\sum_{-90°}^{90°} |E_{ij}(\theta)|^2}{\sum_{-90°}^{90°} |E_0(\theta)|^2} \tag{S28}$$

where subscripts $i$, $j$ can be $+$ or $-$ representing the RHCP and LHCP respectively, $E_{ij}(\theta)$ is the amplitude of the measured electric field scattered from metasurface at an angle $\theta$, while $E_0(\theta)$ is the far-field pattern of the transmitting antenna without metasurface, which act as the reference for the efficiency evaluation.

The transmission efficiency within 8 GHz - 12 GHz of the antisymmetric refractor in Fig. 4A is shown in Fig. S11, where co-polarized transmission efficiency $T_{++}$ and $T_{--}$ are pretty similar and mostly over 0.6 within 8.2 GHz -11.7 GHz, where the maximum is around 0.93 at 10.9 GHz, but drop a bit within 9.2 GHz - 9.6 GHz. Meanwhile, cross-polarized transmission efficiency $T_{+-}$ and $T_{-+}$ remain predominantly below 0.1, indicating that cross-polarized transmission is effectively suppressed and thus demonstrating the dominance of co-polarized transmission. As for the asymmetric refractor in Fig. 4D, its $T_{++}$ and $T_{--}$ are slightly different within 8 GHz - 9.5 GHz and mostly over 0.6 within 8.5 GHz - 12 GHz, and the maximum is around 0.88 at 10.8 GHz, as presented in Fig. S12. These results confirm that co-polarized transmission remains the dominant response. Apparently, the experimental results demonstrate that both metasurfaces maintain high co-polarized transmission efficiency from 8 GHz to 12 GHz, thereby validating the superior performance of the proposed metasurface designs.

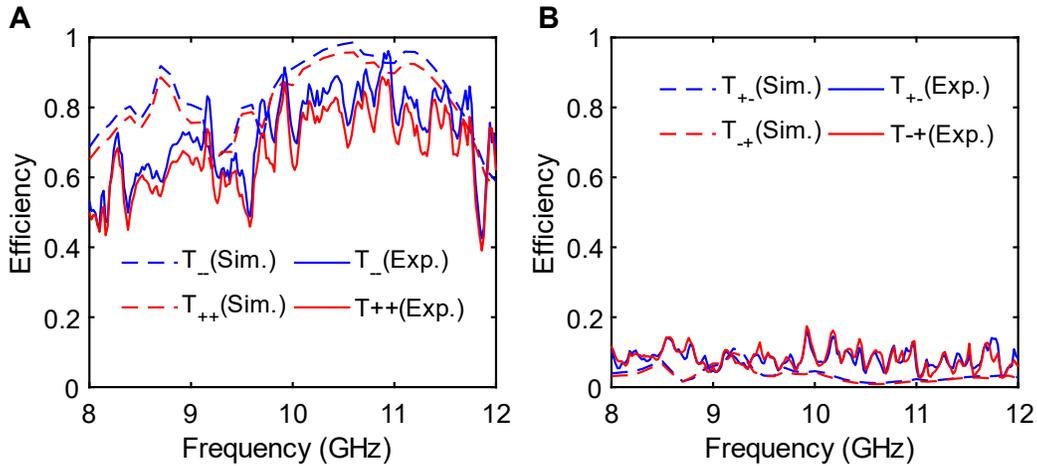

**Fig. S11. The transmission efficiency of antisymmetric refractor (corresponding to metasurface in Fig. 4A in the main text).** The experimental (solid) and measured (dashed) transmission efficiency of (A) co- and (B) cross-polarized channels under RHCP (red) and LHCP (blue) incident wave.

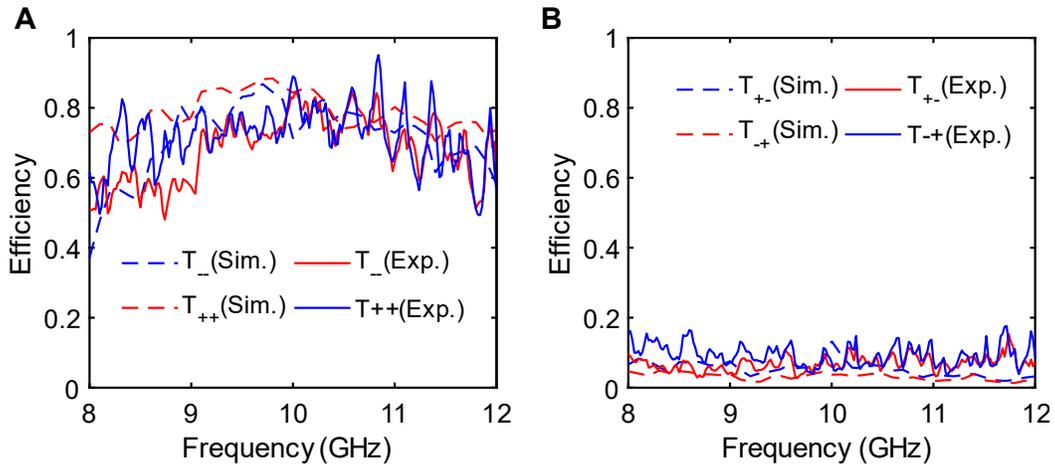

**Fig. S12. The transmission efficiency of asymmetric refractor (corresponding to metasurface in Fig. 4D in the main text).** The experimental (solid) and measured (dashed) transmission efficiency of (A) co- and (B) cross-polarized channels under RHCP (red) and (LHCP) incident wave.

Despite certain fluctuations in the experimental results, there is still a good agreement with the simulation results. The discrepancies between simulations and experiments can be attributed to many factors such as fabrication errors of the prototype, imperfect uniformity of the horn antenna's radiation pattern, phase instability in the vector network analyzer during measurement, and alignment inaccuracies of the devices.

## Supplementary Note 6: Decoupling of two-dimensional wavefront

Besides the asymmetric refraction metasurface with decoupled one-dimension phase profile in the main text, here, a metasurface with decoupled two-dimensional phase profile is proposed and further demonstrated. The target functionalities are focusing and vortex wave generation for RHCP and LHCP incident waves, respectively, and their phase profiles ($\Phi_{++}$ and $\Phi_{--}$), presented in Fig. S13A are

$$\Phi_{++} = \frac{2\pi f_c}{c}\left(\sqrt{x^2 + y^2 + F^2} - F\right) \tag{S29}$$

$$\Phi_{--} = l \cdot \mathrm{atan2}(y, x) \tag{S30}$$

where $x$ and $y$ are the central coordinates of each meta-atom, $F$ is the focal length for focusing, and $l$ is the topological charge of orbital angular momentum (OAM) of the vortex wave. Here, $f_c = 10$ GHz, $F = 150$ mm and $l = 1$, and the design of the metasurface is shown in Fig. S13B.

Fig. S13C and E show the normalized co- and cross-polarized near-field $E_{++}$ and $E_{-+}$ under RHCP illumination, $E_{++}$ presents a clear focusing property on the *yoz* plane, where the focal center is around $z = 100$ mm and the field in the corresponding *xoy* plane show a clear focusing spot. As for $E_{-+}$, the intensity is much darker, proving that cross-polarized transmission is almost absent and thus co-polarized transmission is dominant. Similarly, under LHCP illumination (Figs. S13D and F), $E_{--}$ is a split beam in the *yoz* plane and presents a hollow ring on the *xoy* plane, showing a clear vortex beam property, and we can see a counterclockwise spiral phase from the phase distribution in the inset, proving the topological charge of OAM is 1. The above results prove that the metasurface performs distinct 2D functionalities in the two orthogonal co-polarized transmission channels, further validating the proposed strategy for wavefront decoupling.

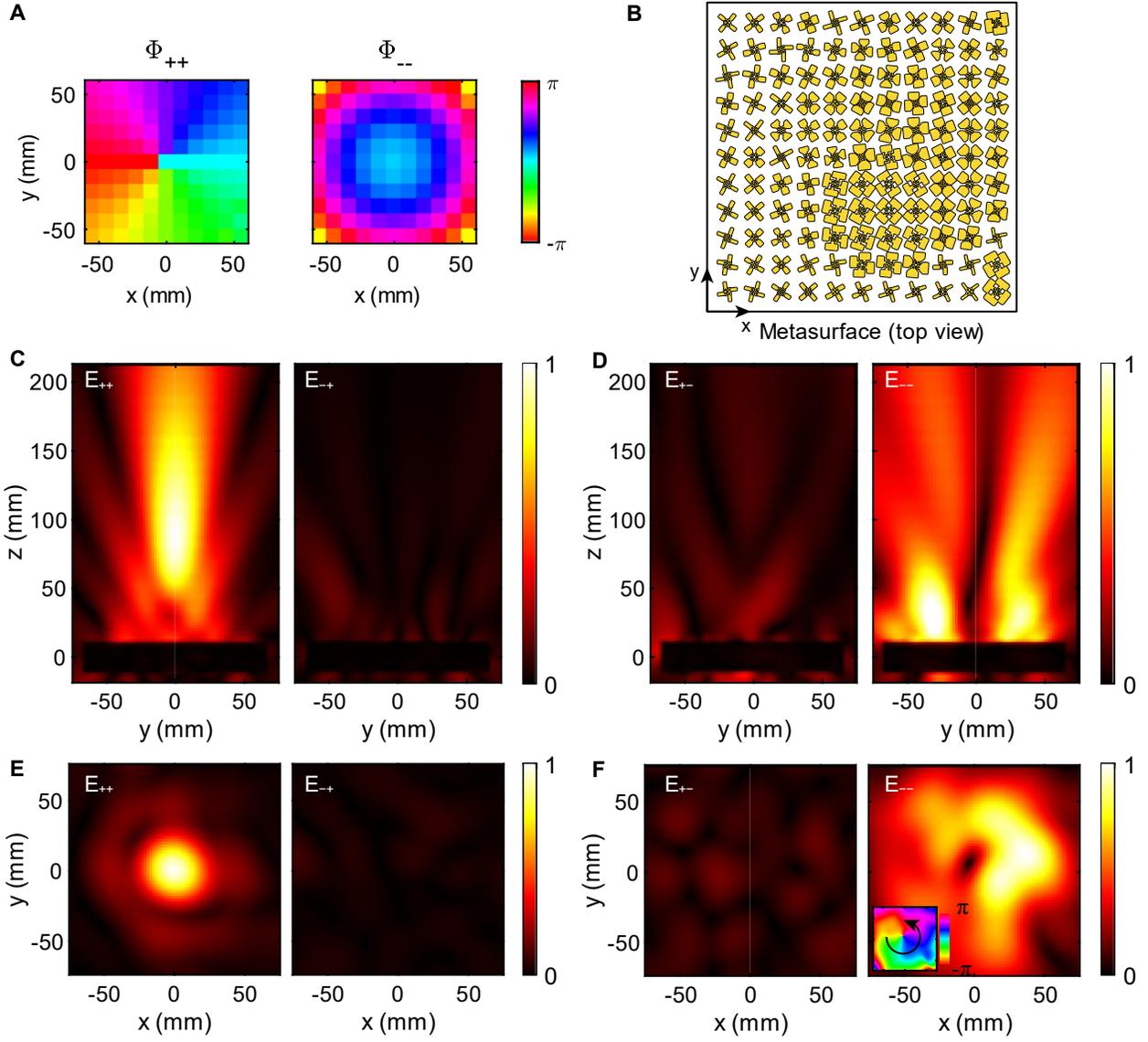

**Fig. S13. Result of the two-dimensional decoupled metasurface.** (**A**) The phase profile $\Phi_{++}(x, y)$ and $\Phi_{--}(x, y)$ of the decoupled metasurface. (**B**) The top view of the metasurface. (**C** and **D**) The normalized near-field of the co- and cross-polarized components on the *yoz* plane under the illumination of RHCP and LHCP wave (10 GHz). (**E** and **F**) The normalized near-field of the co- and cross-polarized components on the *xoy* plane at z = 100 *mm*, and the inset is the phase of E—.

## Supplementary Note 7: Singularity under generic polarization state

The polarization basis of the Jones matrix in Eq. S3 is determined by the polarization at the poles of the sphere, which are orthogonal CP states. Therefore, if the polarization at the poles is replaced by a pair of orthonormal bases $|\psi_+\rangle$ and $|\psi_-\rangle$, the co-polarized channels are under this polarization basis and can be analyzed in the same way. The new Poincaré sphere whose poles are $|\psi_+\rangle$ and $|\psi_-\rangle$ can be obtained by rotating the original sphere around the origin. For simplicity, the cartesian coordinate of original and new spheres is denoted as $\mathbf{S} = (S_1, S_2, S_3)^{\mathrm{T}}$ and $\mathbf{S}' = (S_1', S_2', S_3')^{\mathrm{T}}$, respectively, their relationship is

$$\mathbf{S}' = \mathbf{R} \times \mathbf{S} \tag{S31}$$

and the matrix $\mathbf{R}$ is the three-dimensional rotation matrix follows

$$\mathbf{R} = \begin{pmatrix} \hat{u}_x' \\ \hat{u}_y' \\ \hat{u}_z' \end{pmatrix} = \begin{pmatrix} u_{x,x}' & u_{y,x}' & u_{z,x}' \\ u_{x,y}' & u_{y,y}' & u_{z,y}' \\ u_{x,z}' & u_{y,z}' & u_{z,z}' \end{pmatrix} \tag{S32}$$

$\hat{u}_x' = (u_{x,x}', u_{y,x}', u_{z,x}')$, $\hat{u}_x'$ is the unit $x$-vector on coordinate $\mathbf{S}$' and it is projected to the original coordinate $\mathbf{S}$. $u_{x,x}'$, $u_{y,x}'$, $u_{z,x}'$ are the projection components of $u_x'$ on the original $x$-, $y$- and $z$-axis, where the first subscript denotes the projection component of coordinate $\mathbf{S}$, and the second subscript denotes the component of coordinate $\mathbf{S}$'.

To obtain the matrix $\mathbf{R}$, it is necessary to calculate $\hat{u}_x'$, $\hat{u}_y'$, $\hat{u}_z'$. First, the state of the new $z$-axis ($S_3$-axis) is $|\psi_+\rangle$, the unit vector $\hat{u}_z'$ is its Stokes vector. Second, the state of the new $x$-axis ($S_1$-axis) is $\frac{\sqrt{2}}{2}|\psi_+\rangle + \frac{\sqrt{2}}{2}|\psi_-\rangle$ and its Stokes vector is $\hat{u}_x'$. Third, the new $y$-axis ($S_2$-axis) can be obtained by the cross-product of $u_z'$ and $u_x'$, i.e. $u_y' = u_z' \times u_x'$. Here are their analytical math expressions

$$\hat{u}_x' = \begin{pmatrix} -\sin 2\psi \\ \cos 2\psi \\ 0 \end{pmatrix}, \hat{u}_y' = \begin{pmatrix} -\sin 2\chi \cos 2\psi \\ -\sin 2\chi \sin 2\psi \\ \cos 2\chi \end{pmatrix}, \hat{u}_z' = \begin{pmatrix} \cos 2\chi \cos 2\psi \\ \cos 2\chi \sin 2\psi \\ \sin 2\chi \end{pmatrix} \tag{S33}$$

where $\chi$ and $\psi$ are the parameters of original sphere, and the rotation matrix $\mathbf{R}$ is

$$\mathbf{R} = \begin{pmatrix} -\sin 2\psi & \cos 2\psi & 0 \\ -\sin 2\chi \cos 2\psi & -\sin 2\chi \sin 2\psi & \cos 2\chi \\ \cos 2\chi \cos 2\psi & \cos 2\chi \sin 2\psi & \sin 2\chi \end{pmatrix} \tag{S34}$$

Thus, the transformation relation between $\mathbf{S}$ and $\mathbf{S}$' is

$$\mathbf{S}' = \mathbf{R} \times \mathbf{S} = \begin{pmatrix} S_2 \cos 2\psi - S_1 \sin 2\psi \\ S_3 \cos 2\chi - \sin 2\chi (S_1 \cos 2\psi + S_2 \sin 2\psi) \\ \cos 2\chi (S_1 \cos 2\psi + S_2 \sin 2\psi) + S_3 \sin 2\chi \end{pmatrix} = \begin{pmatrix} \cos 2\chi' \cos 2\psi' \\ \cos 2\chi' \sin 2\psi' \\ \sin 2\chi' \end{pmatrix} \tag{S35}$$

where $\psi'$ and $\chi'$ are the new azimuth angle and ellipticity, respectively. Then the singularity is shown on the surface of the Poincaré sphere in $\mathbf{S}$ and $\mathbf{S}$' coordinates, respectively, for comparison. The singularity in the original sphere (Fig. S14A), the original and new axes are shown parallel, and the singularity is at the antipole of the $S_3'$ axis. By rotating the new axes and sphere simultaneously, as shown in Fig. S14B, the new sphere is similar to the original one, and it can be expanded in the same way (Fig. S14C). Besides, the Jones matrix under orthonormal bases $|\psi_+\rangle$ and $|\psi_-\rangle$ is denoted as $\hat{U}'$ and can be expressed by coordinate $\mathbf{S}$' as

$$\hat{U}' = \exp\left(i\frac{\delta}{2}\mathbf{S}' \cdot \hat{\boldsymbol{\sigma}}^* + i\phi\right) \tag{36}$$

The form is the same as Eq. 1 in the main text, but the polarization base is different. Therefore, the singularity of arbitrary co-polarized channel can be found on the surface of the Poincaré sphere in the same way, proving the generality of the proposed phase mechanism.

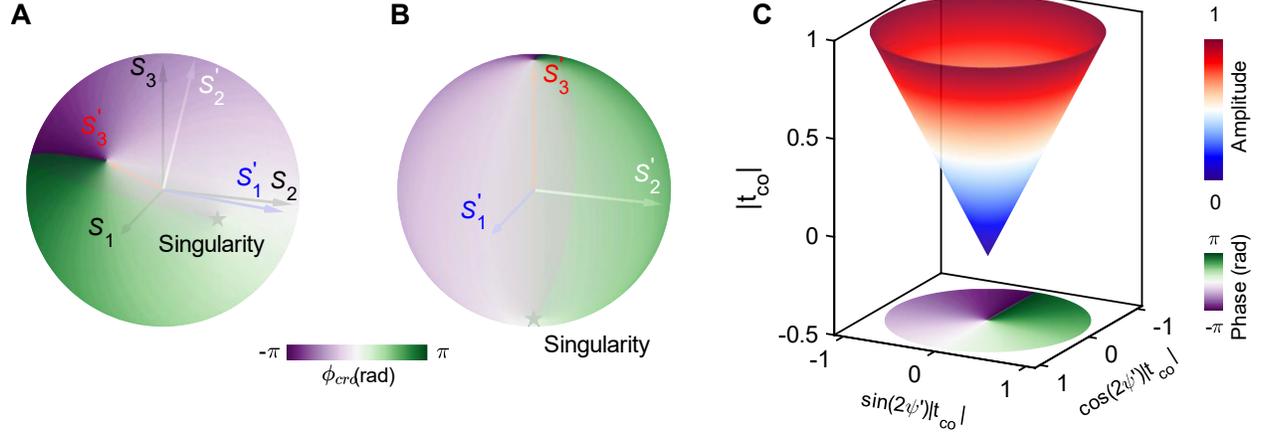

**Fig. S14 The singularity on the surface sphere under arbitrary polarized co-channel.** (A) The singularity and phase on the sphere under arbitrary polarization $|\boldsymbol{\psi}_+\rangle$. Stokes vector of $|\boldsymbol{\psi}_+\rangle$ is the $\boldsymbol{S'_3}$ axis and its antipole is the singularity. (B) Rotate the sphere and the new axis to align with the original axis orientation, the singularity is at the new south pole. (C) Expansion of the rotated sphere.